\documentclass[rmp]{revtex4}
\usepackage{amssymb,amsfonts,amsmath}
\usepackage{graphicx}
\usepackage{amsmath}
\usepackage{natbib}

\newcommand{\EQ}[1]{Eq.~(\ref{eq:#1})}
\newcommand{\EQS}[2]{Eqs.~(\ref{eq:#1}) and (\ref{eq:#2})}
\newcommand{\FIG}[1]{Fig.~\ref{fig:#1}}

\newcommand{\locus}{{\bf s}}
\newcommand{\gt}{g}

\newcommand{\mfit}{\langle F\rangle}
\newcommand{\mafit}{\langle A\rangle}
\newcommand{\mefit}{\langle E\rangle}
\newcommand{\rE}{\tilde{r}}

\begin{document}
\title{Emergence of clones in sexual populations}
\author{Richard~A.~Neher${}^{*}$}
\author{Marija Vucelja${}^{**}$}
\author{Mark~Mezard${}^{\dagger}$}
\author{Boris~I.~Shraiman${}^{\ddagger}$}
\affiliation{${}^{*}$Max Planck Institute for Developmental Biology, 72076 T\"ubingen, Germany}
\affiliation{${}^{**}$Courant Institute for Mathematical Sciences, New York University, New York, NY10012}
\affiliation{${}^{\dagger}$Laboratoire de Physique Th\'eorique et Mod\`eles Statistiques, CNRS and Univ.  Paris-Sud, 91405 Orsay,  France}
\affiliation{${}^{\ddagger}$Kavli Institute for Theoretical Physics and Department of Physics, University of California,
Santa Barbara, CA 91306}
\date{\today}

\begin{abstract}
In sexual population, recombination reshuffles genetic variation and produces novel combinations of existing alleles, while selection amplifies the fittest genotypes in the population. If recombination is more rapid than selection, populations consist of a diverse mixture of many genotypes, as is observed in many populations. In the opposite regime, which is realized for example in the facultatively sexual populations that outcross in only a fraction of reproductive cycles, selection can amplify individual genotypes into large clones. Such clones emerge when the fitness advantage of some of the genotypes is large enough that they grow to a significant fraction of the population despite being broken down by recombination. The occurrence of this ``clonal condensation" depends, in addition to the outcrossing rate, on the heritability of fitness. Clonal condensation leads to a strong genetic heterogeneity of the population which is not adequately described by traditional population genetics measures, such as Linkage Disequilibrium. Here we point out the similarity between clonal condensation and the freezing transition in the Random Energy Model of spin glasses. Guided by this analogy we explicitly calculate the probability, $Y$, that two individuals are genetically identical as a function of the key parameters of the model. While $Y$ is the analog of the spin-glass order parameter, it is also closely related to rate of coalescence in population genetics: Two individuals that are part of the same clone have a recent common ancestor. 
\end{abstract}
\maketitle
Genetic diversity is the fodder for natural selection and the fuel of evolution. It is generated by mutations and by recombination, which reshuffles genomes  and thereby accelerates the exploration of the  space of genotypes. The latter consists of all of the $2^L$ possible combinations of the genetic variants, a.k.a.~alleles, present at $L$ (biallelic) polymorphic loci. Of course the number of polymorphic loci, $L$, itself changes as new mutations arise forming new polymorphisms while ``older" polymorphisms disappear from the population. The population itself consists of $N$ individuals which sample only a small fraction of the possible genotypes, i.e.~$N\ll 2^L$. The dynamics in genotype space is therefore highly stochastic. 

Of particular importance are those genetic polymorphisms that affect the fitness of individuals,  the fitness being defined as the expected number of offspring in the next generation. Selection on its own would amplify the number of high fitness individuals and condense the population into a few ``clones" comprising a large fraction of the population. In populations of sexually reproducing organisms, the growth of such clones and the subsequent decline of genetic diversity are checked by recombination. Two parents, if chosen from different clones, produce offspring that are distinct from either parent. In obligate sexually reproducing species, the formation of clones is prevented since reproduction is coupled to recombination and no parent can produce genetically identical offspring. 
Many species, in particular microbial species and plants, can reproduce both by clonal reproduction (e.g., budding in yeast, selfing or vegetative reproduction in plants) or by sexual propagation. Such facultatively sexual species display a great variety in their mode of propagation, the frequency $r$ of outcrossing, and the heritability of fitness. The latter is very important in sexual populations, as it determines to what extend recombinant offspring benefit from the same fitness advantages that made their parents successful.

The aim of this article is to describe quantitatively the competing tendencies of natural selection and recombination with regard to genetic diversity, focusing on facultatively sexual organisms. The competition of natural selection with recombination is the dominant mechanism of evolution on relatively short time scales, on which mutational input is negligible compared to diversification by recombination. This situation is particularly relevant to adaptation following a major outcrossing event, or within a so called hybrid zone, where diverged genotypes have come together to generate a hybrid population. As this hybrid population continues to breed within itself, it can give rise to a bout of rapid adaptation, as beneficial alleles from both original populations are combined to form novel fit genotypes that spread within the hybrid population \citep{OrtizBarrientos:2002p24186}.

We will focus on the probability, $Y$, that two random individuals sampled from the population have the same genotype, i.e., are clones of each other. This quantity is important for population genetics, since it characterizes genetic diversity (its inverse is a measure of the number of dominant clones) as well as the dynamics of coalescence. Whenever two individuals are part of the same clone, they share a recent common ancestor, such that $Y$ is proportional to the rate of pair coalescence. In the canonical theory of neutral coalescence, this rate is equal to the inverse population size. We will find here that $Y$, and with it the rate of coalescence, is determined by the clonal structure rather than the population size at low outcrossing rates. 

Much of our analysis is presented in the context of a facultatively sexual population, where the evolving entities are individuals and their genotypes. Note, however, that some of our considerations also hold for contiguous segments of chromosomal DNA that are short enough to undergo only infrequent recombination even in obligatory sexual reproduction. In that case, we would be interested in the probability of a given chromosomal segment to be identical for a random pair of individuals drawn from the population. In this context, $Y$ is the homozygosity of the population at this extended locus at which many different alleles segregate (this is of course a much weaker condition than clonal relation of whole genomes). A complementary view of this probability relates it to the ``haplotype diversity", i.e., the number and population distribution of distinct genomic sequences for the chromosomal segment in question. We shall return to this important question in the discussion. 

In an earlier publication \citep{Neher:2009p22302}, we have shown that clones are absent in the so called Quasi Linkage Equilibrium (QLE) ``phase" corresponding to frequent outcrossing limit but appear in a regime of small out-crossing frequency $r<r_c$, with $r_c$ depending on the complexity of the fitness landscape (i.e.~the extent of fitness additivity) and weakly dependent on $N$. We will review this finding and present a more detailed analysis of the time dependence of this condensation phenomenon, as well as its quantitative dependence on fitness additivity and hence heritability. Furthermore, we also study the extend of clonal condensation in a steady state where the fitness distribution is moving towards higher fitness at a constant velocity. 
We will put these results into the context of the Random Energy Model (REM) of Statistical Physics, introduced and solved by Bernard Derrida  \citep{Derrida:1981p30865,Gross:1984p29586,Mezard:2009p28797}. In fact, $Y$ is closely related to the Parisi order parameter and the onset of clonality is closely related to the spin-glass transition observed in simple models of disordered media such as the REM.

Below, we will first draw the analogy between the dynamics of selection in finite populations and the REM. This analogy is particularly simple for $r=0$. Whereas the condensation transition in the REM occurs below a certain critical temperature, the transition to clonal population structure occurs beyond a certain critical time,  $t > t_c (N)$. Hence the population genetic analog of temperature will be the inverse time. We shall then generalize the model in order to include (facultative) recombination and fitness landscapes with varying degrees of epistasis, i.e., genetic interactions. The results for mixed epistatic and additive models enables us to set the analysis into the context of the ``traveling wave" approximation, which has recently emerged as a powerful representation of adaptive population dynamics in genetically diverse populations \citep{Tsimring:1996p19688,Rouzine:2003p33590,Cohen:2005p45154,Desai:2007p954,Neher:2010p30641,Hallatschek:2011p39697}. Our results therefore provide insight into how recombination and epistasis affect the dynamics and structure of adapting population waves and define conditions under which genetic diversity is maintained or lost.

\section{Natural selection and the Random Energy Model.}

In the absence of recombination or mutation, the frequency of any individual
genotype increases if its fitness lies above the population mean and decreases
otherwise. Identifying fitness with relative growth rate and ignoring stochastic
effects, the expected number of individuals $n_i(t)$ with genotype $g_i$ and
fitness $F_i$ obeys:
\begin{equation}
\dot{ n}_i (t)= (F_i-\mfit_t)n_i(t)  \quad \Rightarrow \quad n_i(t) = e^{F_i t-
\int_0^t dt' \mfit_{t'} } \ ,
\end{equation}
where $\dot{ n}_i(t) $ is the time derivative of $n_i(t)$ (we assume $n_i(0) = 1$) and the mean fitness  $\mfit_t$ is defined by averaging over the whole population. Defining the rate of growth of clones by differential fitness (relative to the population mean) ensures constant population size: $\sum_i n_i (t) =N$.
Since the mean fitness term is shared by every genotype in the population, the
frequency of a particular genotype is given by 
\begin{equation}
\nu_i(t) = \frac{e^{F_i t}}{\sum_j e^{F_j t}} = Z(t)^{-1}e^{F_i t} \ ,
\end{equation}
where $Z(t)$ is a time dependent normalization constant known as partition function.
Hence the distribution of clones in the population has the Boltzmann form, where
inverse time plays the role of temperature. The dynamics of the population will
now depend on the initial distribution of fitness values $F_i$. In the generic
case where fitness depends on many loci, each one giving a small contributions
to fitness, the density of fitnesses $\rho(F)=\mathcal{N}(0,\sigma^2)$ is
approximately Gaussian (with zero average and variance $\sigma^2$) and the
statistics of clones in the populations is identical to that of the REM 
\citep{Derrida:1981p30865}. For small $t$ population averages are dominated 
by the vicinity of the peak of $\rho (F)$, with many
individual genotypes contributing. However, for $t>t_c$, the dominant
contribution shifts all the way to the leading edge 
$F_{max}\approx\sigma \sqrt{2\log N}$, which corresponds to the maximum fitness sampled
from the Gaussian $\rho (F)$ in a population of size $N$. This means that with
increasing time, i.e., decreasing ``effective temperature", the population shifts
to fitter and fitter genotypes and eventually, for $t>t_c$, ``condenses" into the
fittest. This condensation phenomenon manifests itself in a non-negligible
probability $Y$ that two randomly chosen individuals from the population have
the same genotype. The latter is equal to the average squared genotype frequency 
in the population at time $t$. The average participation ratio is obtained by averaging 
over the fitness values $\{F_i\}$ of the $N$ initial genotypes
\begin{equation}
\label{eq:YREM}
\langle Y_t\rangle =\left \langle \sum_i \left(\frac{n_i(t)}{\sum_j n_j}\right)^{2} \right \rangle_{\{F_i\}}= 
N\int_0^\infty dz\; z \int dF_i \rho(F_i) n_i^2(t) e^{-z n_i(t)} \left[\int dF_j \rho(F_j) e^{-z n_j(t)}\right]^{N-1} \ ,
\end{equation}
where we have used the integral representation of $\Omega^{-2} = \int_0^\infty dz\ ze^{-z\Omega}$ and the fact
that the $F_i$ are the i.i.d.~random variables sampled from $\rho(F)$. This calculation is carried out
explicitly in \citep{MPV:1985,Mezard:2009p28797} and leads to the following 
quantitative result: In the limit of large populations, $\langle Y_t\rangle$ is
given by
\begin{equation}
\label{eq:YRESULT}
\langle Y_t\rangle=\begin{cases}
	\mathcal{O}(N^{-1})\quad t<t_c\\
	1-\frac{t_c}{t}\quad t>t_c
\end{cases} \ ,
\end{equation}
with $t_c \approx \sigma^{-1} \sqrt{2\log N}$. \FIG{REM} shows how $\langle Y_t\rangle$, measured in a computer simulation (see below), increases in time and compares to the REM prediction. Note that the sharp transition exhibited in \ref{eq:YRESULT} is realized only in the limit of large $\log N$, which is hard to achieve both in reality and in a numerical simulation. In both  cases one expects to find a crossover rather than a sharp transition. Nevertheless, considering the idealized $N \rightarrow \infty$ limit provides a very useful scaffold for the analysis, and the REM also allows us to compute the $1/\log N$ corrections and therefore determine the detailed nature of the crossover.
\begin{figure}[htp]
\begin{center}
  \includegraphics[width=0.5\columnwidth]{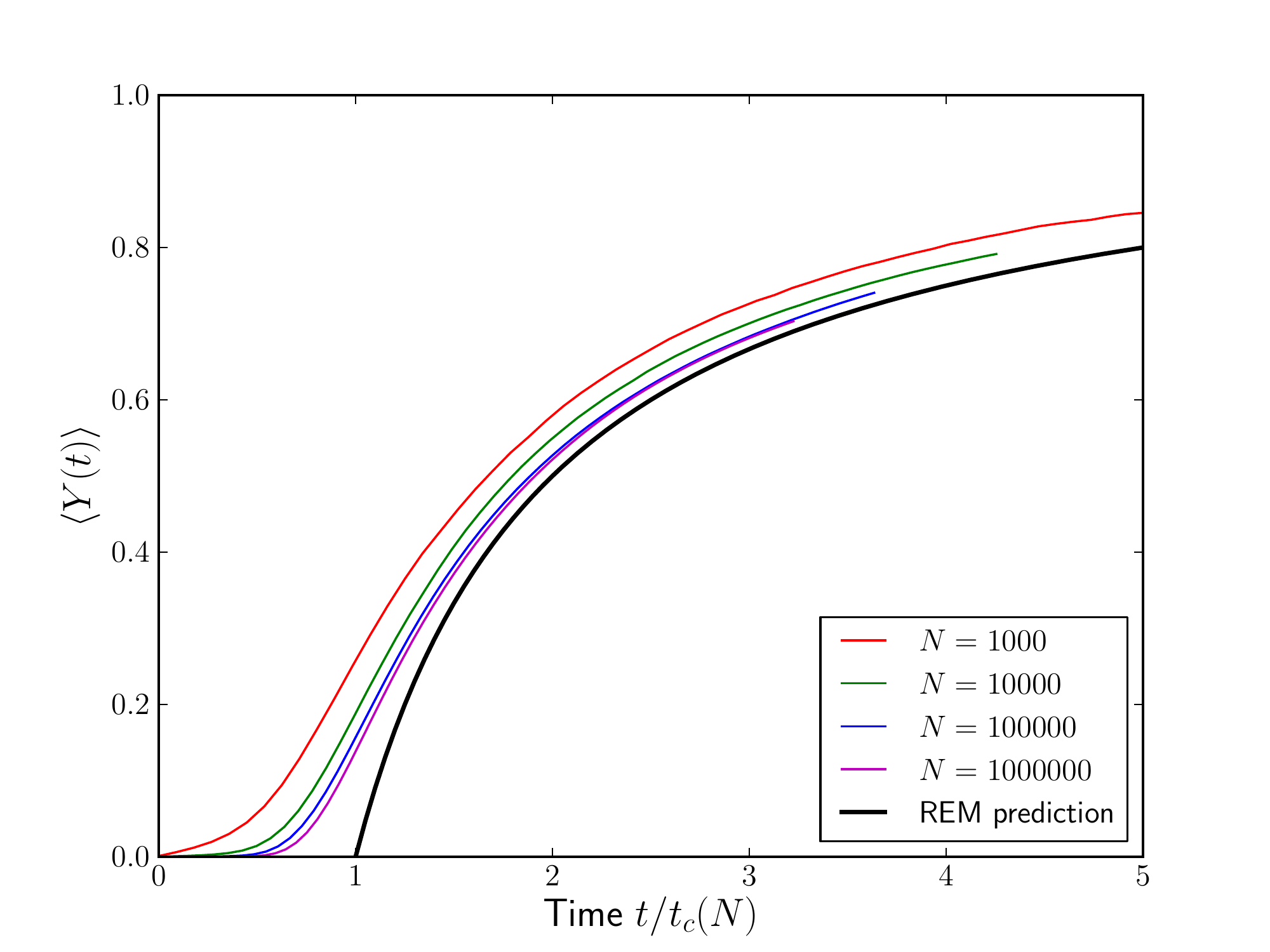}
  \caption[labelInTOC]{The participation fraction $\langle Y_t \rangle$ in the absence of recombination as a function of time. For each $N$, the time axis is rescaled by $t_c = \sigma^{-1}\sqrt{2\log N\sigma}$. The fitness variance $\sigma^2 = 0.0025$ and is included in the expression for $t_c$ to account for stochastic effects. The solid black line shows the $N\to\infty$ asymptotic result predicted by the REM (\EQ{YRESULT}). The convergence of the numerical results to this asymptotic result  with increasing $N$ is slow because it is governed by $\log N$. Each curve is averaged over 1000 runs. }
  \label{fig:REM}
\end{center}
\end{figure}

\subsection*{Recombination and heritability}

Sexual reproduction mixes the genetic material from two parents and thereby
produces new genotypes from existing genetic variation. To account for
recombination, we modify the equation describing the evolution of genotypes as
follows:
\begin{equation}
\label{eq:hypercube_evo}
\dot n(\gt) = (F(\gt)-\mfit)n(\gt) + r \left[N^{-1}\sum_{\gt', \gt''} K(\gt|\gt', \gt'')n(\gt')n(\gt'') -n(\gt)\right]
\end{equation}
where $n(\gt)$ is the number of individuals with genome $\gt$, and $K(\gt|\gt',
\gt'')$ accounts for the probability that genotype $\gt$ is assembled by
recombination from the parental genotypes $\gt'$ and $\gt''$.
In the absence of recombination, the only relevant characteristic of a genotype
is its fitness, and we could study the evolution of the population along the fitness coordinate
instead of on the hypercube of possible genotypes. To achieve a similar simplification 
with recombination we need to know how the fitness of recombinant offspring relates to
that of the parents to map \EQ{hypercube_evo}. The correlation between parental and offspring traits is
known as heritability and depends on the underlying genetic architecture. If a
trait depends on many loci in the genome in an additive manner, i.e., different
loci make independent contributions to the trait, the trait value of offspring
will be approximately Gaussian distributed around the parental mean. Such traits
are called highly heritable since the correlation between trait values of
parents and offspring is high. Conversely, if a trait depends on specific
combinations of alleles at many loci (epistasis), these combinations will be
disrupted with high probability in sexual reproduction.
Such traits have a low heritability in sexual reproduction since the
correlation between parents and offspring is low.

\begin{figure}[htp]
\begin{center}
  \includegraphics[width=0.7\columnwidth]{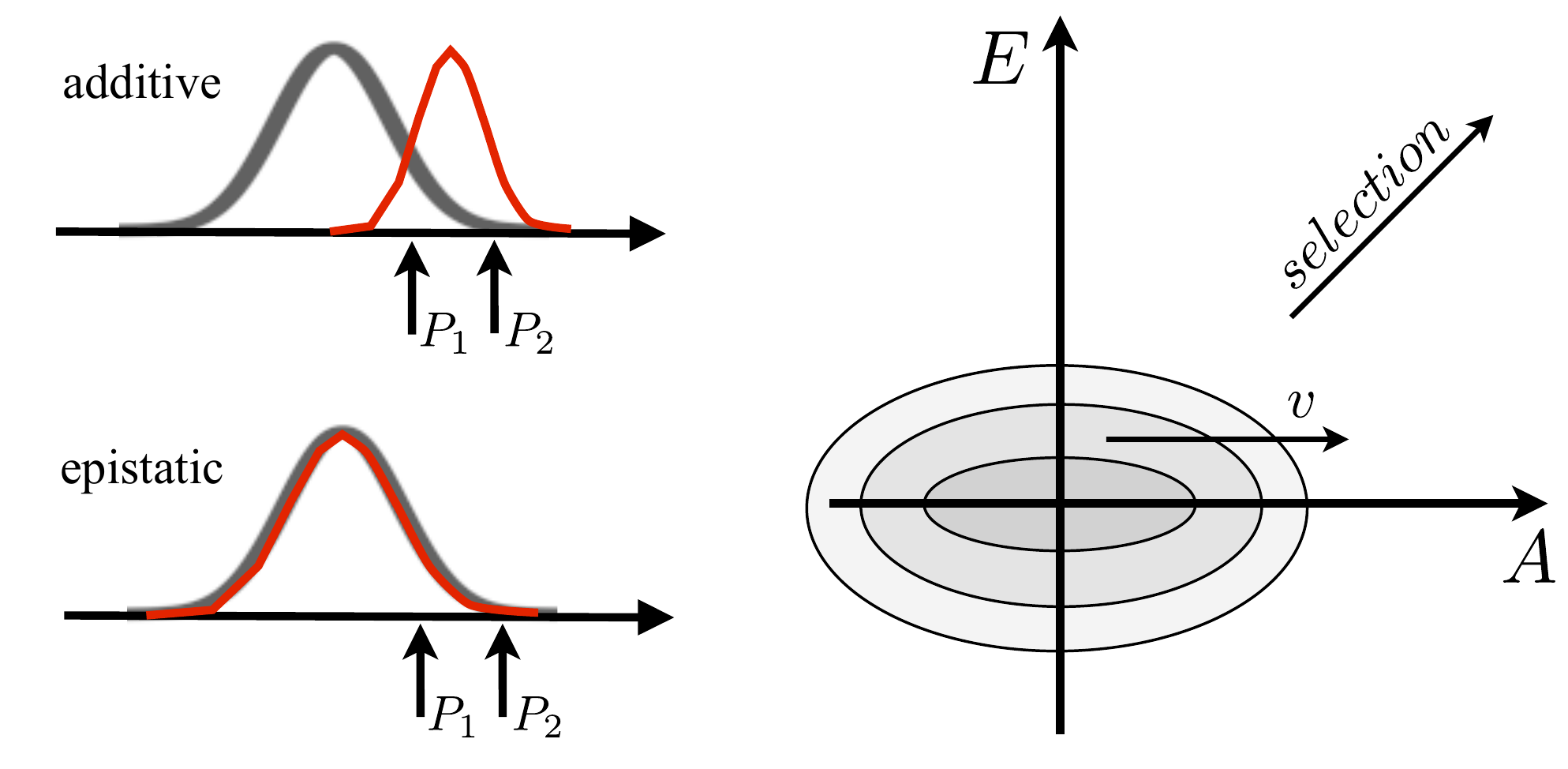}
  \caption[labelInTOC]{
Heritable and non-heritable contributions to fitness. Left: The two panels illustrate heritable and non-heritable fitness functions. The black Gaussian represents the density of states over all $2^L$ possible genotypes, while the arrows indicate the fitness of parents $P_{1,2}$ sampled from this density of states. For additive fitness functions (top), the fitness distribution of recombinant off-spring of parents $P_{1,2}$ (red curve) is centered around the mid-parent value, i.e., fitness is heritable. For completely epistatic fitness functions (bottom), the offspring fitness is a random sample from the density of states and therefore not heritable. Right: Fitness is a sum of additive (heritable) and epistatic (non-heritable) components and selection amplifies individuals with large $F=A+E$. In sexual reproduction, only the additive component of fitness is heritable, so only the additive fitness increases in time with rate $v$.
}
  \label{fig:heritability}
\end{center}
\end{figure}

The trait we are mainly interested in here is fitness, which in general involves
many phenotypes and depends on the environment. We shall set aside the issues
associated with fluctuating environment and possible time-dependence of
selection, and focus on how fitness depends on the genotype. As already noted,
the genotype space is a high dimensional hypercube, $\gt=(s_1,\dots,s_L)$, and
for present purposes, the fitness is a complicated (but fixed)  function of the
genotype parameterized as
\begin{equation}
F(\gt) = f_0 + \sum_i f_i \locus_i + \sum_{i<j} f_{ij} \locus_i\locus_j + \cdots \ ,
\end{equation}
where the terms $f_i \locus_i$ define the additive contribution of locus $i$ to
fitness, while higher order terms correspond to epistatic interactions. 
The relation of the fitness of an offspring relative to
that of its parents and the density of states, i.e., the distribution of fitness
over all possible genomes, is illustrated in \FIG{heritability}.

The higher the order of the interactions, the less likely it is that a
particular set of loci that contributes to one parent is inherited uninterupted.
As a consequence, the heritability of interaction terms goes down with
increasing order. Interaction terms of high-order are essentially independent of
the parents. This is reminiscent of high-order spin glass models, where the
energies of any two configurations that differ at a macroscopic number of spins
are uncorrelated. The large $p$ limit of such $p$-spin glasses is the random
energy model, where the energy of each configuration is an independent draw from
a Gaussian distribution \citep{Derrida:1981p30865,Gross:1984p29586}.

\section{The Model}
In general the genetic architecture of fitness is expected to be complex with
additive contributions as well as epistatic contributions of various orders. It
is very instructive, though, to consider a simplified model that includes a
heritable component and a random epistatic component. Specifically,
let us assume that fitness can be decomposed into an additive component $A$ and an
epistatic component $E$. If two individuals with fitness $F_1=A_1+E_1$ and
$F_2=A_2+E_2$ produce an offspring,  its additive component, $A$, is drawn from a
Gaussian with variance $\sigma_A^2/2$ around the parental mean $(A_1+A_2)/2$,
while its epistatic component, $E$, is independent from that of the parents and is drawn from a
Gaussian centered around $0$ with variance $\sigma_E^2$. Hence the recombination kernel of this  model depends only on the additive fitness of the parents 
\begin{equation}
\label{eq:rec_kernel}
K(A,E|A_1,A_2) = \frac{1}{\sqrt{2}\pi \sigma_E \sigma_A}e^{-\frac{(A-(A_1+A_2)/2)^2}{\sigma_A^2}-\frac{E^2}{2\sigma_E^2}}
\end{equation}
For much of the analysis below, we will simplify this model even further and
assume that the additive fitness of recombinant genotypes is
independent of that of the parents and simply drawn from a Gaussian with
variance $\sigma_A^2$ centered around the population mean $\mafit$. This distribution is expected if the distribution of $A$ in the population is approximately Gaussian. This simplified model is easier to analyze, while displaying the same qualitative behavior, as has been checked by numerical simulations. The joint distribution of $A$ and $E$ in the population then evolves according to 
\begin{equation}
\dot P(A,E) = (F-\mfit -r)P(A,E) + \frac{r}{2\pi \sigma_E \sigma_A}e^{-\frac{(A-\mafit)^2}{2\sigma_A^2}-\frac{E^2}{2\sigma_E^2}}
\end{equation}

The ratio of $\sigma_A$ and $\sigma_E$ define the extent of fitness heritability in the process of recombination. We define ``heritability", $h$, which will be one of the key independent parameters of the model, as: 
\begin{equation}
h^2=\sigma_A^2/(\sigma_A^2+\sigma_E^2)
\end{equation}

To illustrate the behavior of the model at different recombination rates and different heritabilities, we implemented it as a computer simulation. The computer program keeps track of clones with fitness $A_i$ and $E_i$  and population size $n_i$. At each generation, the size of a clone is updated by a Poisson distributed number with 
mean  $n_i e^{F_i-\mfit-r+C}$, 
where $C = (1-N/N_0)$ is a density regulating term. A clone is deleted if its size is 0. In addition, at each generation, $Nr$ new clones of size 1 are seeded, with additive fitness drawn from a Gaussian $\mathcal{N}(\mafit,\sigma_A)$ and epistatic fitness drawn from $\mathcal{N}(0,\sigma_E)$. Alternatively, we can impose a ``velocity'' of the additive fitness by drawing its value from  $\mathcal{N}(vt,\sigma_A)$ instead of  $\mathcal{N}(\mafit,\sigma_A)$.

Due to the stochastic nature of reproduction, the majority of all initial genotypes will rapidly die out, even if very fit. Of the $N$ initial genotypes, only a fraction $\sim \sigma$ remains, where $\sigma^2 = \sigma_A^2+\sigma_E^2$ is a measure of the typical strength of selection. Similarly, of the clones produced by recombination, only a fraction $\sigma$ ``establishes". Those clones that do establish are on average larger than the deterministic expectation by a factor of $\sigma^{-1}$. We will neglect them in most of the formulae below. The dominant effect of this stochasticity can be accounted for by rescaling $N$ to $\sigma N$ inside logarithms. We will reinstantiate this correction in comparisons to simulations when necessary. This stochastic effects are of minor importance for the phenomena we study here since they are overshadowed by the randomness inherent in the fitness and seeding time of new clones.

\section{Results: Structure of adapting populations}
Depending on the rate of recombination $r$ and the degree of fitness heritability $h$, the model formulated above exhibits very different behaviors. Before we present a formal characterization of the population structure and the dynamics of evolution, it is instructive to discuss the dynamics of the model as observed in simulations. 

\begin{figure}[htp]
\begin{center}
  \includegraphics[width=0.48\columnwidth,type=pdf,ext=.pdf,read=.pdf]{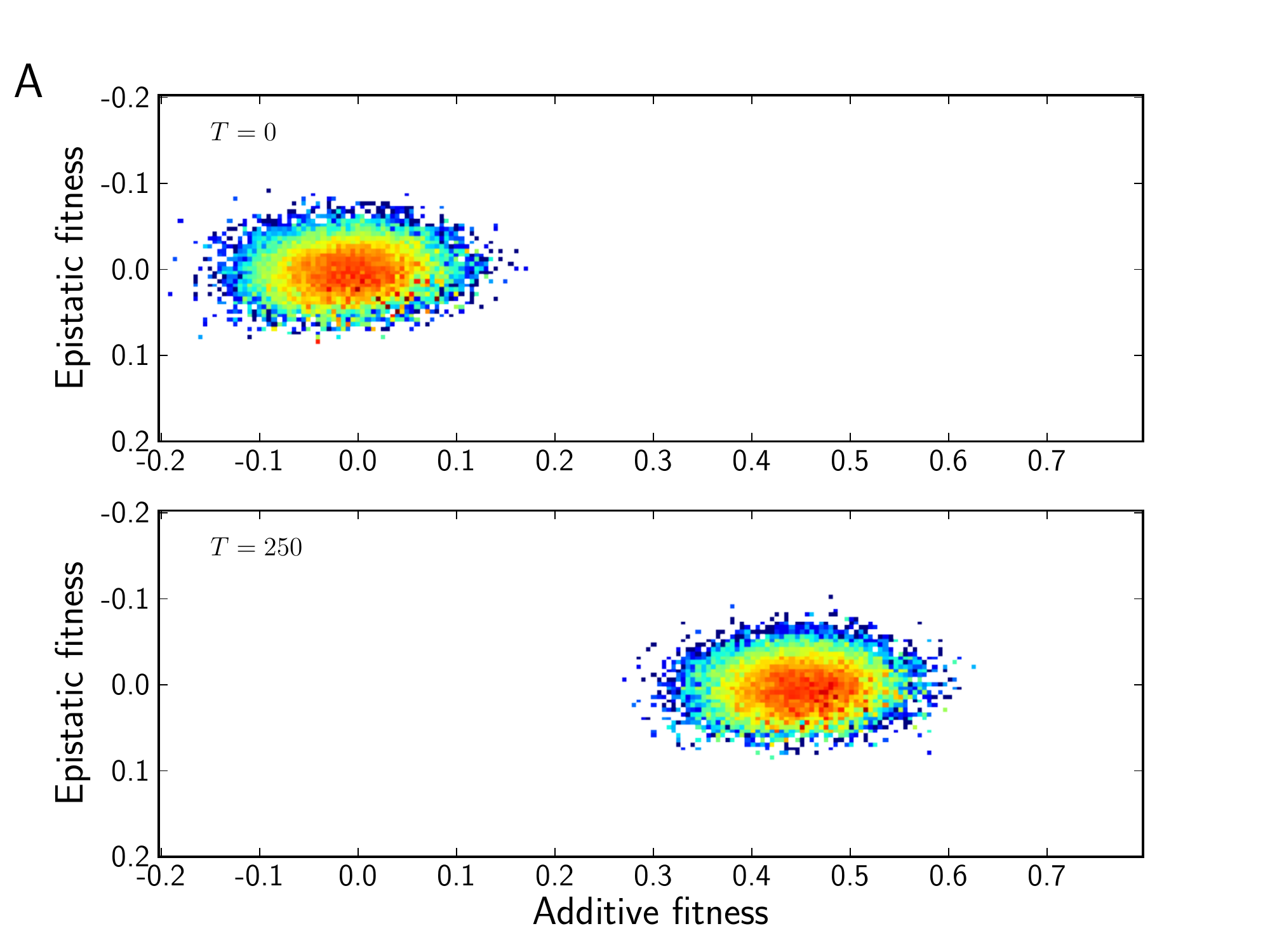}
  \includegraphics[width=0.48\columnwidth,type=pdf,ext=.pdf,read=.pdf]{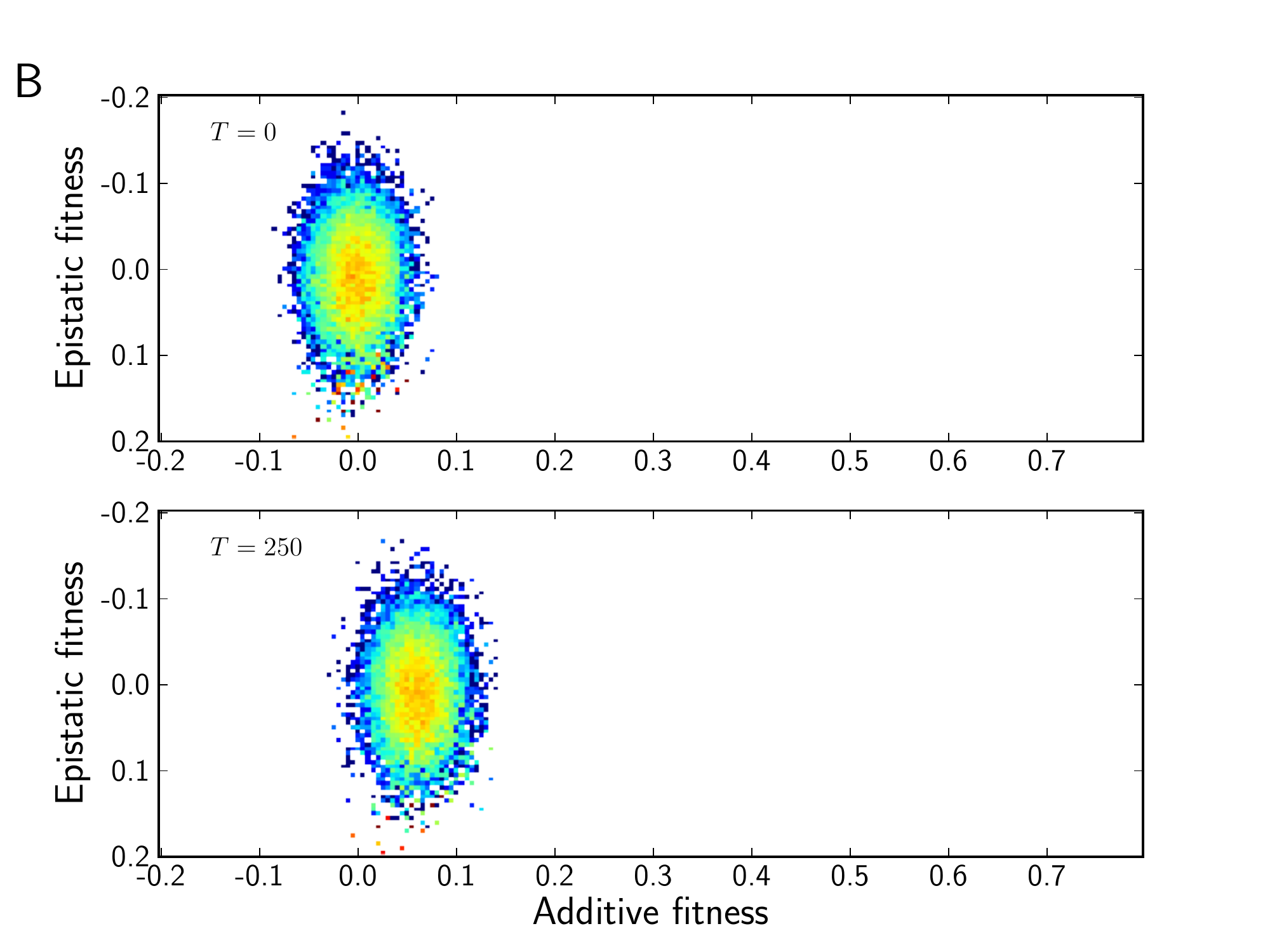}
  \caption[labelInTOC]{The distribution and dynamics of additive and epistatic fitness in the population at moderate recombination rates $r=\sigma$. Panel A: Additive fitness dominates ($\sigma_A^2=0.8\sigma^2$, $\sigma_E^2=0.2\sigma^2$). Panel B: Epistatic fitness dominates ($\sigma_A^2=0.2\sigma^2$, $\sigma_E^2=0.8\sigma^2$). In both panels, $\sigma^2=0.0025$ and $N=10^{5}$.}
  \label{fig:movie}
\end{center}
\end{figure}

\FIG{movie} shows the distribution of additive and epistatic fitness at two
different times for scenarios where additive (left) or epistatic (right) fitness
dominates.
If fitness is predominantly additive, the population adapts and moves towards
high additive fitness as long a new genotypes are generated by recombination.
The velocity is given by the variance of the population along the additive
direction. If most of the variance is along the epistatic direction as in the right
panel, the adaptation of the population is much slower. In addition, large
fractions of the population tend to be condensed into a small number of clones,
as we will discuss at greater length below. As the variance along the additive
direction decreases to zero, so does the velocity.

\begin{figure}[htp]
\begin{center}
  \includegraphics[width=0.98\columnwidth]{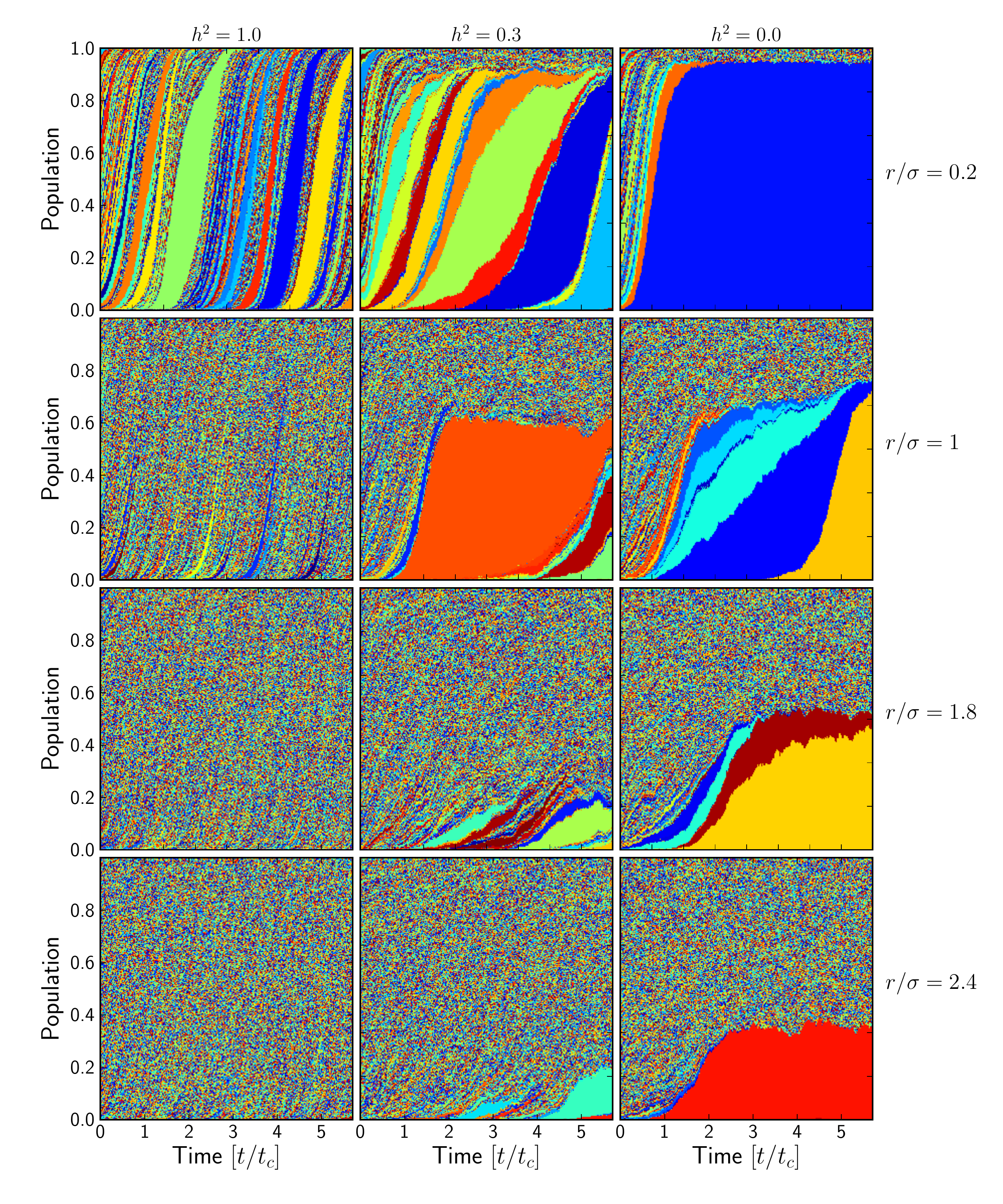}
  \caption[labelInTOC]{Structure of population for different recombination rates and heritabilities; see text for discussion. In all panels, $\sigma^2=0.0025$ and $N = 10^{4}$.}
  \label{fig:clones}
\end{center}
\end{figure}

The population structure and dynamics at different recombination rates are
illustrated in \FIG{clones}. The figure shows the composition of the population
as a function of time for different recombination rates and different
heritabilities. In each panel, the population was initialized as a diverse
sample from the density of states and was subsequently allowed to evolve via
selection and recombination. Hence each panel shows a transient, which gives way
to a steady state behavior. Each genotype in the population is assigned a
specific color, and individuals are ordered according to fitness in each time
slice with the fittest individuals at the bottom.

The left column of \FIG{clones} shows evolution governed by an all additive
fitness function ($\sigma_A=\sigma$, $\sigma_E=0$) for different recombination
rates. In this case, the population moves steadily to higher fitness since new
genotypes, fitter than their parents, are constantly produced. At low
recombination rates, the population consists of a small number of clones that
arise at the high fitness edge (bottom part of the panel), grow as time
progresses, and shrink again once they fall behind in fitness (i.e., when they
move towards the center of the panel). As the recombination rate is increased to
values comparable to $\sigma$, large clones cease to exist.
Most of the population is made from nearly unique genotypes that are short
lived. This pattern becomes even more pronounced at larger recombination rates.
The high recombination limit of this dynamics can be understood in mean field
theory (MFT) described in Section \ref{sec:MFT}. The low recombination limit is
described in greater detail in Section \ref{sec:v_less_one}.

The right column of \FIG{clones} shows the opposite limit, when fitness is not
heritable but completely epistatic, corresponding to a fitness function with
random high order interactions. In this case, we do not expect the population to
move towards higher fitness indefinitely since the epistatic fitness is
non-heritable. At low recombination rates, we expect that the fittest genotype
in the population will grow while generating new recombinants distributed around
zero fitness. With the growing genotypes, the mean epistatic fitness will
increase until the selection on the fittest genotype, $E_{max}-\mefit$, equals
the rate at which it produces recombinants. This behavior is clearly seen in all
four panels with $h^2=0$ on the right. Since the recombination rate increases
from top to bottom, the size at which the fittest genotype stabilizes decreases,
while the ``dust"-like fraction of the population that consists of short-lived
unfit recombinants  increases. Occasionally, recombination generates exceptionally fit genotypes, 
which have a chance of replacing the previous record
holder. As time progresses, these records become rarer and rarer according to
the well known results on record dynamics\footnote{Note that a slightly
different dynamics is observed if recombination of individuals with the same
genotype do not produce a novel genotype. In this case the effective outcrossing
rate of large clones goes down and the population quickly condenses into a
unique clone.}. The clone structure and its dynamics for the fully epistatic
case will be discussed in Section \ref{sec:v_zero}. The high recombination limit
can again be understood within mean field theory; see Section \ref{sec:MFT}.

The center column of \FIG{clones} shows the clone structure and dynamics of a
case of intermediate heritability. At low recombination rate, the population is
dominated by clones, and clones exist even when the additive population is only
dust. However, unlike the $h^2=0$ case, no clone dominates forever, but new
clones are continuously established. The velocity in this regime will depend
critically on how often such new clones are produced and on how much they
advance the additive fitness. We will investigate this case below in Section
\ref{sec:v_less_one}.

\subsection{Quasi-Linkage Equilibrium (QLE) and its breakdown}
\label{sec:MFT}
At large $r$, this model admits a factorized solution $P(A,E) =
\vartheta(A,t)\omega(E)$ with 
\begin{equation}
\vartheta(A,t) =
\frac{e^{-\frac{(A-\sigma_At)^2}{2\sigma_A^2}}}{\sqrt{2\pi\sigma_A^2}} \quad\quad \mathrm{and} \quad\quad
\omega(E) =
\frac{r}{r-\mefit-E}\frac{e^{-\frac{E^2}{2\sigma_E^2}}}{\sqrt{2\pi \sigma_E^2}}
\end{equation}
derived in \citep{Neher:2009p22302}. This solution travels toward higher
additive fitness with a velocity equal to the  variance of the additive
fitness of recombinant offspring, while the epistatic fitness has steady
distribution where $\mefit$ adjusts itself to normalize the distribution. This
factorization is a hallmark of Quasi-Linkage Equilibrium
\citep{Kimura:1965p3008, Turelli:1994p2652,Neher:2011p45096}, where additive
components evolve independently, while epistatic components are in a
quasi-steady balance between selection and recombination.
However, this factorized solution breaks down as soon as there are individuals
with epistatic fitness larger than $r+\mefit$. In that case this solution is no
longer normalizable, and additive and epistatic components cease to be
independent.

\begin{figure}[tbp] \includegraphics[width=0.49\columnwidth]{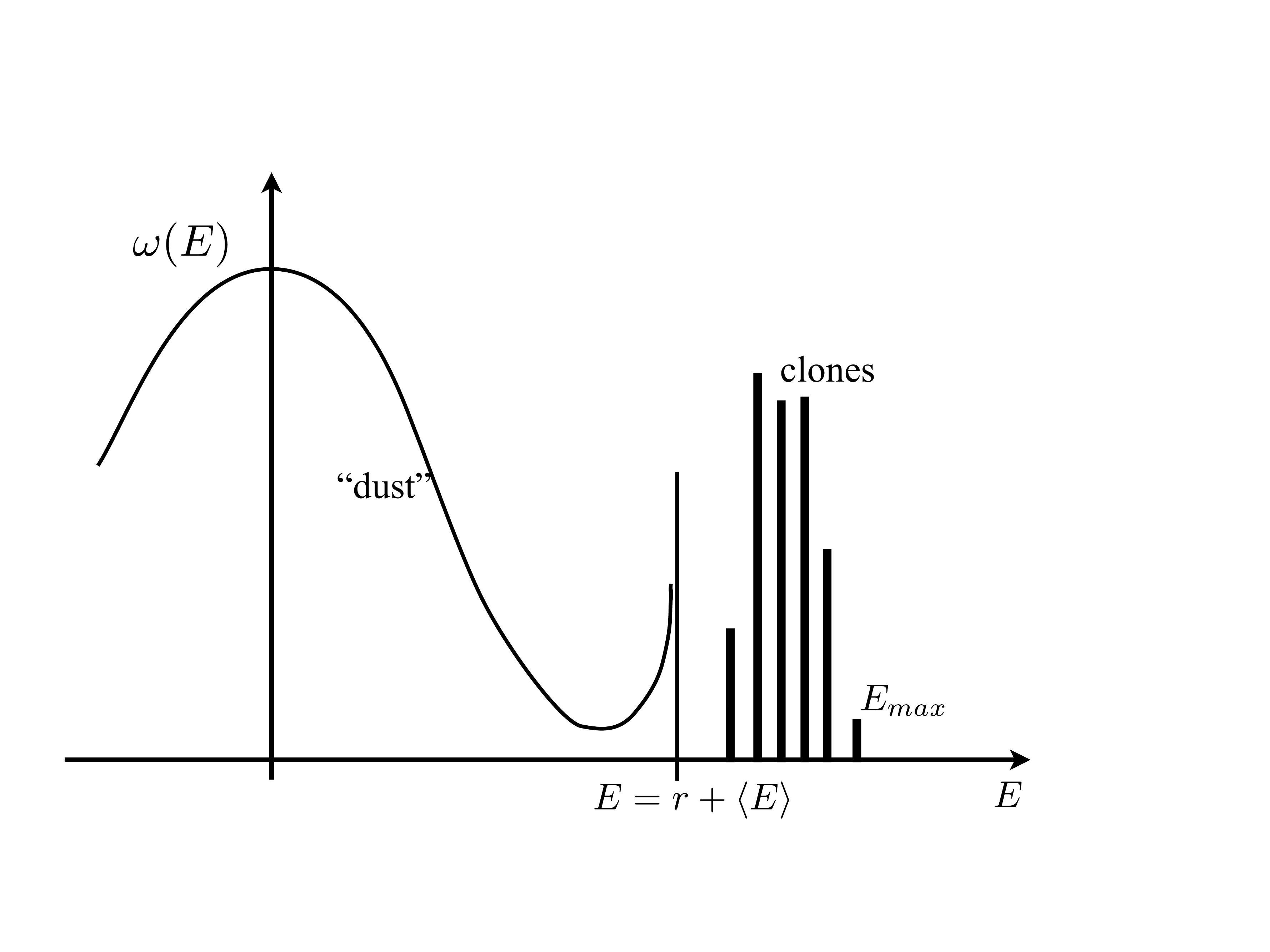}
\caption{\label{fig:back+clones}The population consists of mediocre recombinant
genotypes (dust) and leading clones.  A sketch of the population structure,
indicating the smooth quasi-static distribution $\omega(E)$ at $E<\mefit+r$ and the
exponentially growing clones with $E>\mefit+r$. }
\end{figure}

\FIG{back+clones} illustrates this condensation behavior. The smooth distribution 
$\omega(E)$ is a deformed Gaussian which diverges at $E=r+\mefit$. Beyond
$r+\mefit$, the population consists of growing clones whose size depends on when 
and where they were seeded.

In the following we will characterize the population
structure and study how the velocity in the direction of increasing additive
fitness depends on parameters. We will begin by studying the purely
epistatic case with recombination, which extends the REM by
continuous seeding of novel recombinant genotypes. We will then study the
condensation phenomenon in presence of additive fitness, where the
population forms a traveling pulse in fitness.

\subsection{Clonal condensation for zero heritability}
\label{sec:v_zero}
Suppose we start with a diverse sample of size $N$ from the distribution of
epistatic fitness and inject new recombinant genotypes with rate $Nr$. After a
time $t$, the population consists of $N$ initial clones and a Poisson
distributed number of new clones. A clone of age $\tau_i$ and fitness $F_i=E_i$
will have approximately size
\begin{equation}
n_i(t) = e^{(E_i-r)\tau_i+\int_{t-\tau_i}^t dt' \mefit_{t'}} \ ,
\end{equation}
where the mean fitness $\mefit_{t'}$ can be thought of as a Lagrange parameter that keeps the overall population size constant. To calculate $\langle Y_t(r,h)\rangle$ we have to average over the fitness of the $N$ initial clones and over the fitnesses and seeding times of all subsequently produced recombinant genotypes:
\begin{equation}
\label{eq:def_Y}
\langle Y_t(r,h)\rangle =\int dz \; z\left[NB_0(z) + \sum_{k=1}^{\infty} \frac{e^{-Nrt}}{k!} k B_r(z)\right] [1-C_0(z)]^{N}C_r(z)^{k-1} \ ,
\end{equation}
where $B_0(z)$ is the average of $n_i^2(t)e^{-zn_i(t)}$ over the $E_i$ of initial genotypes  (note that ``initial" means that $\tau_i=t$). $B_r(z)$ is the corresponding average over recombinant clones, which in addition are averaged over their age $\tau_i$; see Appendix \ref{sec:app_definitions} for derivation. Similarly, $C_0(z)$ and $C_r(z)$ are averages of $e^{-z n_i(t)}$ over the $E_i$ and $\tau_i$. The sum over $k$ -- the number of recombinants generated in time $t$ -- can be performed easily. These integrals are evaluated in the Appendix \ref{sec:app_hzero} in the limit of large $N$. One obtains an approximation
\begin{equation}
\langle Y_t(r,0)\rangle  \approx \exp\left[-Ne^{-\sqrt{2\log N}  [1-{r \over r_c }]t\sigma } \right]
\end{equation}
valid for $r \sim r_c$, with
\begin{equation}
r_c \approx \sigma \sqrt{2 \log N} .
\end{equation}
We observe that the participation ratio
substantially deviates from zero only for $r<r_c$ and only upon reaching $t>t_c(r)$, where for $r\approx r_c$
\begin{equation}
\label{eq:tc_prediction}
t_c(r) =\sigma^{-1}{\sqrt{{1 \over 2} \log N} \over 1-{ r \over r_c}} = \frac{t_c(0)}{2}\frac{r_c}{r_c-r} \ .
\end{equation}
This provides us with estimates of the critical time of condensation $t_c(r)$ and the critical $r_c$ below which condensation occurs.
Condensation is delayed compared to the case of $r=0$ and the critical time (analogous to inverse critical temperature) diverges as the recombination rate approaches its critical value $r_c$. Of course, this divergence only occurs in the limit of $\sqrt{\log N}$ going to infinity. In practice, with realistic population sizes $N$, one observes not a transition, but merely a crossover between different regimes of behavior. More elaborate expressions for $t_c$ and $r_c$ at finite $N$ can be found in Appendix \ref{sec:app_hzero}. Note that the values of $t_c$ and $r_c$ themselves scale with the square root of the logarithm of $N$.

\begin{figure}[btp]
\begin{center}
  \includegraphics[width=0.49\columnwidth]{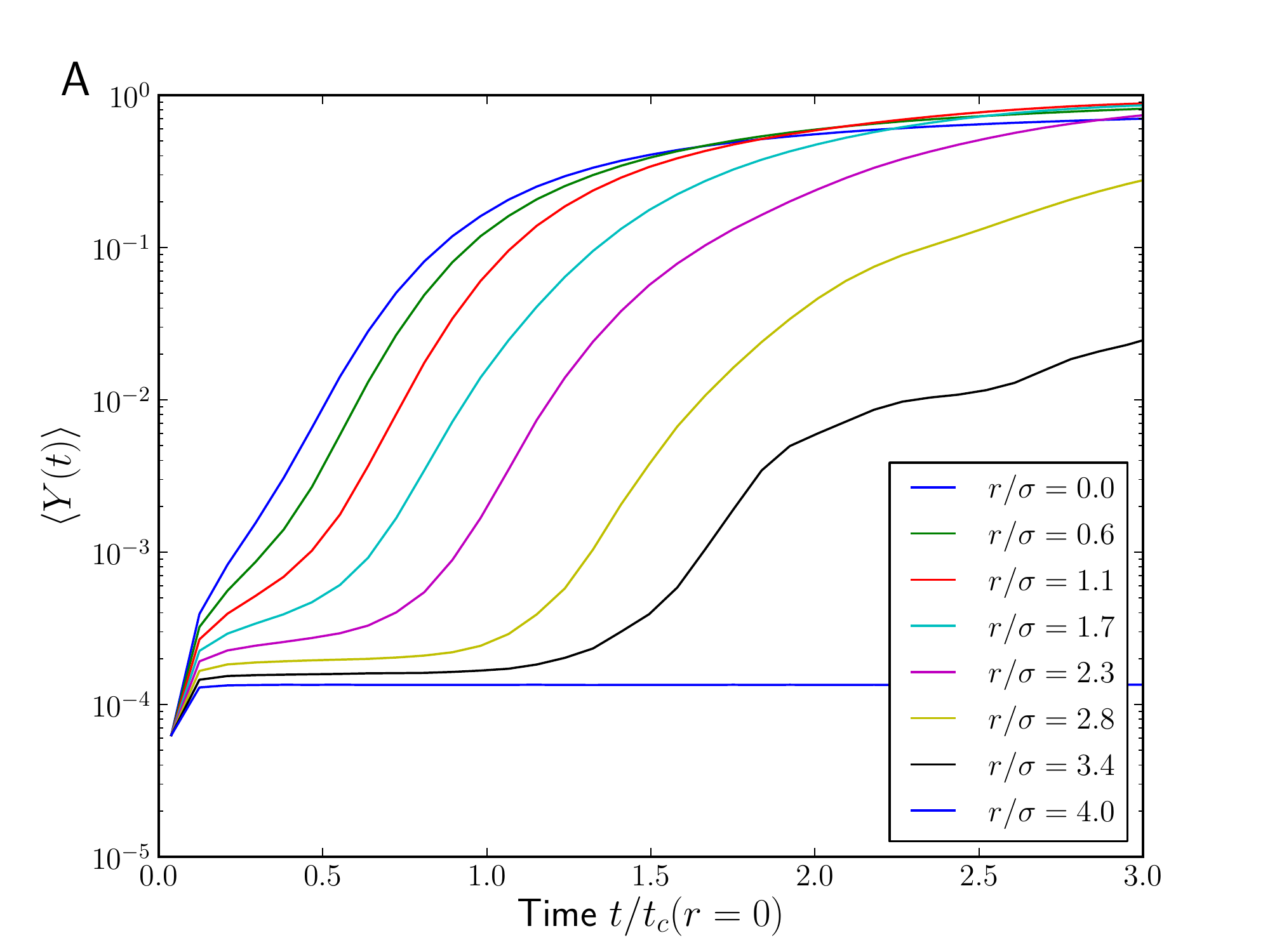}
  \includegraphics[width=0.49\columnwidth]{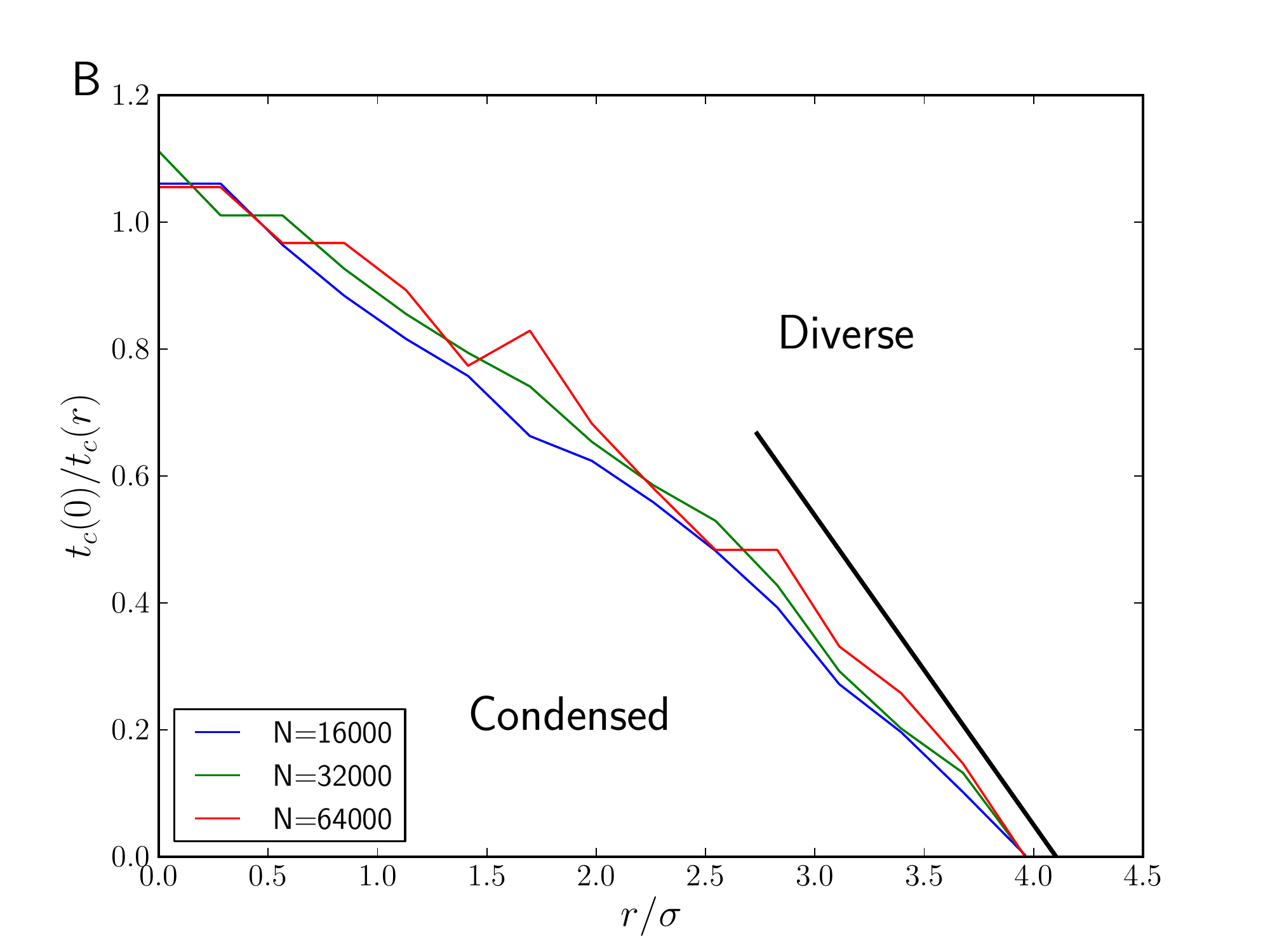}
  \caption[labelInTOC]{
  Panel A shows $\langle Y_t(r,0) \rangle$ for different ratios of $r/\sigma$ for fully epistatic fitness functions as a function of time relative to the critical time with $r=0$. The curves shown are averages over many runs which exhibit substantial stochasticity. Recombination delays condensation and reduces $\langle Y_t(r,0) \rangle$ early on. The model used to produce the data in this plot accounts for the fact that recombination of identical parents does not produce a novel genotype, which results in rapid condensation once one clone becomes macroscopic. Parameters $N=16000$, $\sigma^2=0.005$, and $h^2=0$.
  Panel B shows how the condensation time $t_c(r)$ diverges as $r$ approaches $r_c\approx E_{max}\approx \sigma\sqrt{2\log N\sigma}$. For times and recombination rates below the lines, the population is condensed into clones; above the lines, no genotype is populated by a macroscopic fraction of the population. Here, $t_c(r)$ is defined empirically by $\langle Y_{t_c}(r)\rangle =0.1$. The solid black line indicates the prediction of \EQ{tc_prediction} for $r\approx r_c$. } 
  \label{fig:Y_tc_epistatic}
 \end{center}
\end{figure}

$\langle Y_t(r,0)\rangle$ is shown in \FIG{Y_tc_epistatic}A for different recombination rates. With increasing recombination rate, the early plateau of $\langle Y_t(r,0)\rangle$ is reduced and condensation is delayed.  Each individual run condenses rapidly once the dominant clone is large enough that it often recombines with itself, which leaves the genotype intact. Also, the lack of condensation for $r>r_c$ is clearly seen. \FIG{Y_tc_epistatic}B shows the inverse time to condensation for different recombination rates and population sizes, confirming \EQ{tc_prediction} for $r\approx r_c$.
This transition with increasing $r$ was identified in \citep{Neher:2009p22302}. Intuitively, the divergence of the time to condensation is due to the fact that the growth rate of best genotypes decreases with increasing recombination load $r$. As soon as $E-r$ is smaller than zero for all existing genotypes, all clones are short-lived and no condensation can occur. Similar behavior has been observed in populations with heterozygote advantage or disadvantage \citep{Franklin:1970p13820,Barton:1983p34506}.

Population dynamics for $r<r_c$, including $t>t_c$, has the nature of a ``records process"  \citep{Krug:2005p13598,Sire:2006p29956}. As time progresses, more and more genotypes are sampled from the density of states and tested by selection. As a consequence, the population will come across fitter and fitter genotypes, resulting in a record process with $Nrt$ trials. Even if initially no genotype with $E-r>0$ is present, such a genotype will eventually be found and result in condensation. On the other hand, any finite population will eventually reach a final ``record" (with fitness $E_f$), giving rise to the clone that will eventually take over the whole population. This is because a clone rapidly fixates once it is large enough to frequently recombine with itself. To prevent its fixation a new record would have to be created with the fitness advantage $\Delta E > \tau (E_f-r)^2/\log N$ within the time delay of $\tau$, which is very unlikely beyond certain $E_f$.

\subsection{Traveling solutions for additive fitness}
\label{sec:v_equal_one}
In the opposite limit when $h^2=1$, the population moves towards higher fitness
with a velocity that is given by the additive variance $v=\sigma_A^2$ for
sufficiently large $r$. It will be useful to parameterize the ratio between the velocity 
$v$ and the scale of selection $\sigma^2$ as $\gamma = v/\sigma^2$. At high recombination
rates, we have $\gamma = h^2$, while we generally expect $\gamma<h^2$ at low 
recombination rates. In contrast to the case $h^2=0$, no aging
dynamics is observed for $h^2>0$. Instead, old genotypes are constantly replaced, and
dominant genotypes in the population have a finite characteristic age. Hence the
initial genotypes are rapidly forgotten and we can restrict the analysis to
recombinant genotypes. 

Genotypes seeded at the high fitness nose of the population
distribution can grow large and dominate the participation ratio $\langle Y
\rangle$. Since $\langle Y \rangle$ is closely related to the rate of
coalescence, it is instructive to calculate $\langle Y \rangle$ explicitly assuming
$v=\sigma^2$ before considering the case $v<\sigma^2$ at low $r$ or $h^2<0$.

\begin{figure}[htp]
\begin{center}
  \includegraphics[width=0.48\columnwidth]{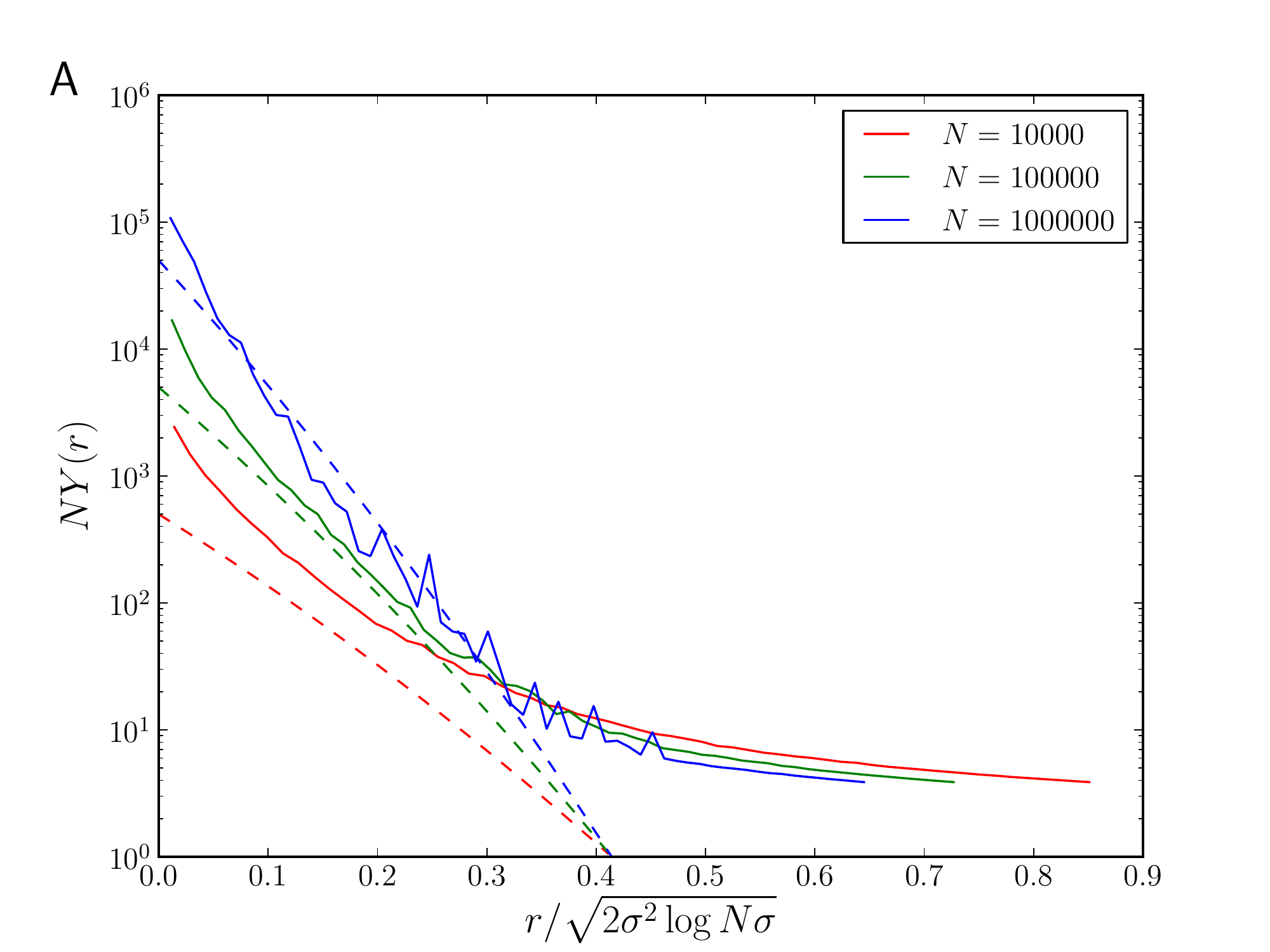}
  \includegraphics[width=0.48\columnwidth]{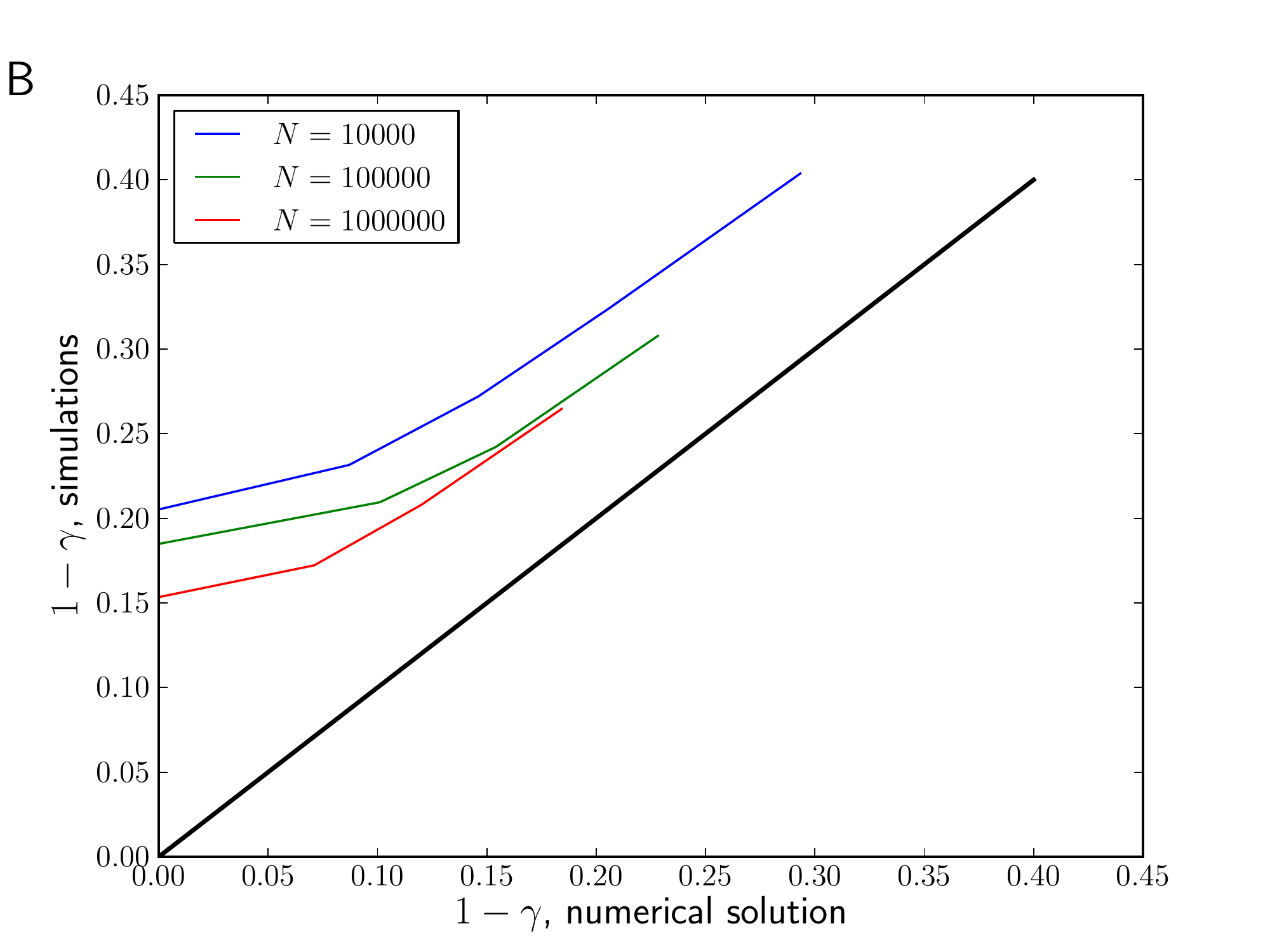}
  \caption[labelInTOC]{Panel A: The rescaled average participation fraction $\langle N Y(r) \rangle$ 
  as a function of $r/\sqrt{2\sigma^2\log N\sigma}$. For large $r$, we find $N\langle Y(r) \rangle\sim 1$, 
  while $N\langle Y(r) \rangle\gg 1$ for small $r$. The dashed lines indicate the prediction by 
  \EQ{Yvone}, which is expected to be valid  for $r<(\sqrt{2}-1)\sqrt{2\sigma^2\log N}$. The theory curves 
  are scaled such that $Y=N^{-1}$ at the crossover. 
   Panel B plots the reduction of velocity observed in simulations against 
   the numerical solution of \EQ{v_additive} where $\gamma=v/\sigma^2$.}
  \label{fig:Y_and_v_additive}
\end{center}
\end{figure}

In the steady state, a clone is specified by its age $\tau_j$ and its initial
fitness $A_j$. Assuming $v=\sigma^2$, we then find
\begin{equation}
n_j = e^{(A_j-r)\tau_j - \frac{v \tau_j^2}{2}} \ .
\end{equation}
To evaluate $\langle Y(r,1)\rangle$, we have to evaluate the integrals $B_r(z)$
and $C_r(z)$ that appear in \EQ{def_Y} and are defined in the Appendix \ref{sec:app_definitions}, see
\EQ{def_Br}. The integrals involve averaging over the initial fitness $A_i$ and
all possible ages $\tau_i$. For $h^2 = \gamma=1$, one finds  $C_r(z)\approx z$ because
relatively young and small clones dominate. 
In contrast, $B_r(z)$ has a significant contribution from rare old
clones, which dominate because of their large size (through the $n_i^2$ factor).
The evaluation of the integrals is detailed in Appendix \ref{sec:Ytraveling} with the result
\EQ{app_Br_traveling}. We obtain
\begin{equation}
B_r(z)
\approx \alpha + z^{-1}e^{-\frac{r}{\sigma}\sqrt{2\log z^{-1}}-\frac{r^2}{2\sigma^2}}\Gamma(-\frac{r}{\sigma\sqrt{2\log z^{-1}}}) \ .
\end{equation}
The first term is of order 1 and corresponds to the contribution of young
clones. Its exact value depends on the details of the stochastic dynamics, such
as the offspring distribution. The second term is the contribution from old large
clones. The participation ratio therefore becomes
\begin{equation}
\label{eq:Yvone}
\langle Y(r,1) \rangle \approx N \int_0^\infty dz \; B_r(z) e^{-Nz} \sim \begin{cases}  \alpha N^{-1} & r/\sigma>(\sqrt{2}-1)\sqrt{2\log N\sigma} \\
e^{-\frac{r}{\sigma}\sqrt{2\log N\sigma}-\frac{r^2}{2\sigma^2}} & r/\sigma<(\sqrt{2}-1)\sqrt{2\log N\sigma} \ .
\end{cases}
\end{equation}
For small $r$, $\langle Y(r,1) \rangle$ does not scale as $N^{-1}$. In other
words, the larger the population is, the larger are the clones it is composed of
and those largest clones dominate $\langle Y \rangle$. This result is in
agreement with arguments made for rapid coalescence in facultatively sexual
populations in \citep{Neher:2011p42539,Rouzine:2010p33121,Rouzine:2007p17401}.
\FIG{Y_and_v_additive}A shows $\langle N Y(r) \rangle$ for as a function of
$r/\sqrt{2\sigma^2\log N\sigma}$. As soon as $r/\sqrt{2\sigma^2\log N\sigma}<0.4$, $Y(r)$ increases and no
longer scales with $N$, as predicted by the mean field calculation above. The
figure also shows the explicit expression in \EQ{Yvone} as dashed lines. Note,
however, that the calculation of $\langle Y(r,1) \rangle$ involved several
approximations where $\sqrt{2\log N}$ was assumed large. As a result, the
accuracy of the prefactor is low, and we have dropped all non-exponential parts,
while enforcing $\langle Y(r,1) \rangle = 1/N$ at the cross-over.

The derivation above is valid for $r\sigma^{-1}\sqrt{2\log N\sigma}>1$. For smaller $r$, the
velocity drops below the high recombination limit, $\gamma=1$, and the fitness
distribution in the population stops being Gaussian. In this regime, the
velocity has to be calculated self-consistently by matching the rate of
production of new extremely fit clones to the overall velocity of the
population. This problem has been studied by
\citet{Rouzine:2005p17398,Rouzine:2010p33121} in the context of HIV evolution.

A simplified version of the arguments of \citet{Rouzine:2005p17398} is given in
Appendix \ref{sec:rouzine}. With $\gamma = v/\sigma^2$, we find
\begin{equation}
\label{eq:v_additive}
1 - \gamma =  \frac{\log [\frac{\sigma}{r}\sqrt{2(1-\gamma)}]}{\log [\frac{\sigma}{r}\sqrt{2(1-\gamma)}]+\log [\gamma^{-1}rN]}
\end{equation}
This implicit equation for $\gamma$ can either be solved numerically or, in the case
of very small $r$, iteratively.
The result is similar to that obtained by \citep{Rouzine:2005p17398};
differences are due to the difference in the model used. The numerical solution is
compared to simulation data in \FIG{Y_and_v_additive}B. Convergence with increasing $N$ 
is slow and \EQ{v_additive} has a solution only for small $r$, for which we have little data.

The velocity of the traveling wave falls below $\sigma^2$ as soon as a few large clones
dominate the population. In this case, not only $B_r(z,h)$ but also $C_r(z,h)$
are dominated by those large clones, and $\langle Y(r,h)\rangle$ behaves
differently. We show in the appendix that $\langle Y(r,h)\rangle \sim 1-\gamma$ is of
order 1 (see \EQ{app_Yvlessone}). This is in agreement with
\citet{Rouzine:2005p17398}, who found that the typical number of large clones is
$\sim \log  r N\sigma \sim Y(r,h)^{-1}$.

\subsection{Traveling wave solutions at intermediate heritabilities}
\label{sec:v_less_one}
In the high recombination limit, the population moves towards higher fitness
with velocity $v=\sigma_A^2$ regardless of any epistatic fitness contributions.
However, we have seen above that a clonal structure can emerge even in the absence
of epistasis if recombination rates are low enough. Intuition and the simulation
in \FIG{clones} show us that this clonal structure is more pronounced in the
presence of epistasis and persists at high recombination rates.

The reason for this behavior is the fact that the fitness variance of the
recombinant offspring is larger than the velocity
$v=\sigma^2_A<\sigma^2$. In this case, fit genotype grow above
the traveling Gaussian envelope and generate macroscopic clones.

\begin{figure}[htp]
\begin{center}
  \includegraphics[width=0.32\columnwidth]{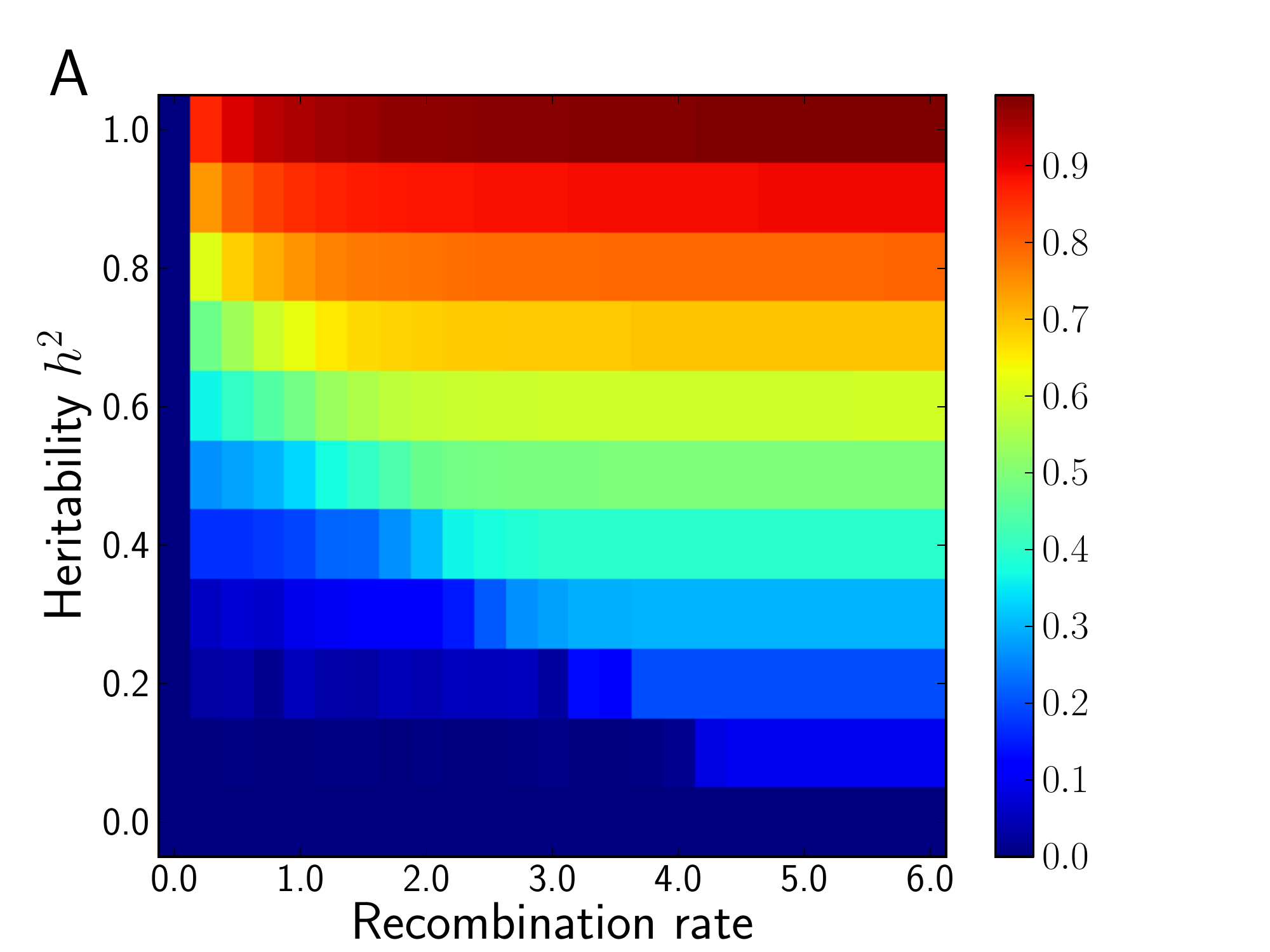}
  \includegraphics[width=0.32\columnwidth]{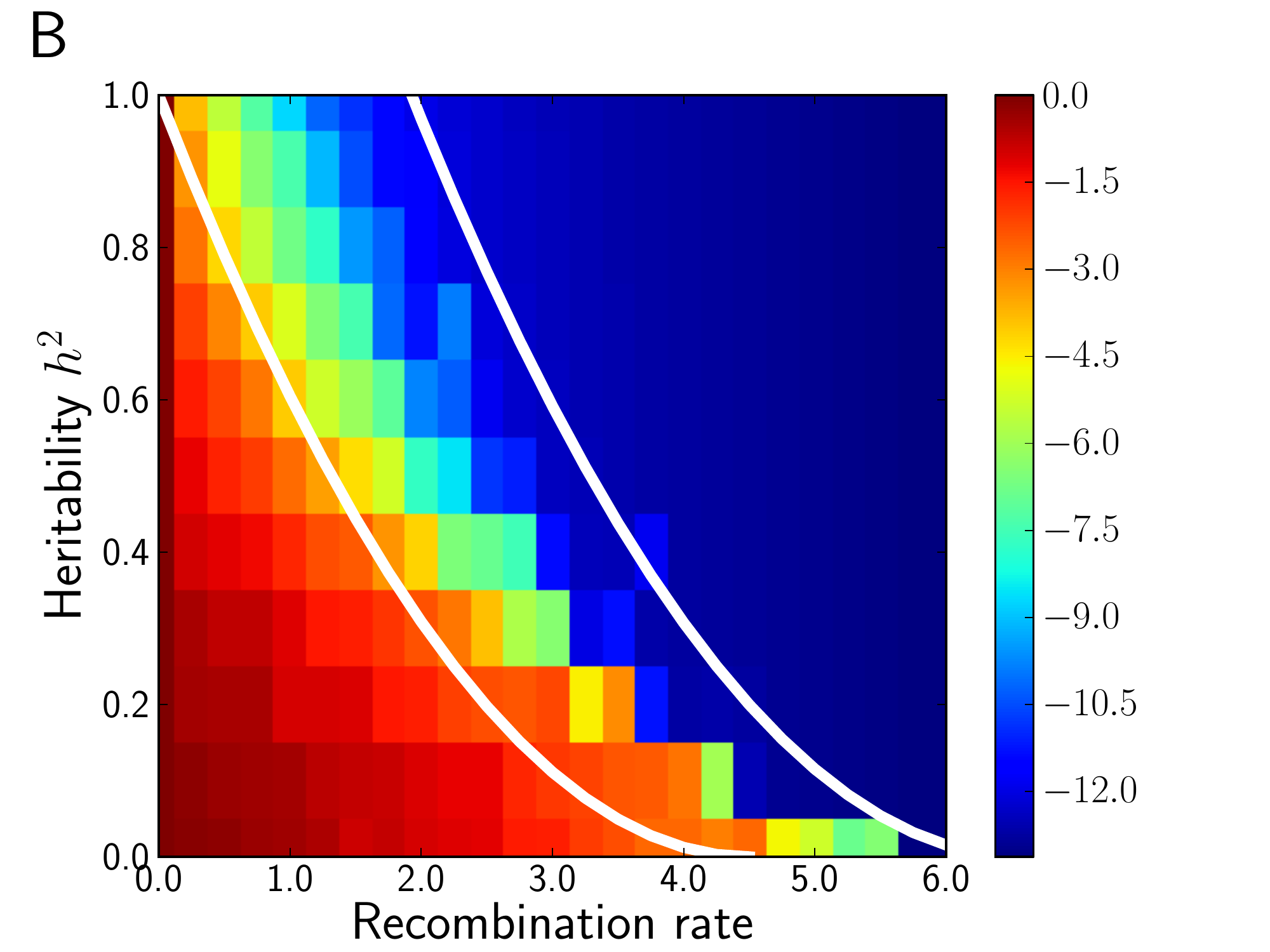}
	\includegraphics[width=0.32\columnwidth]{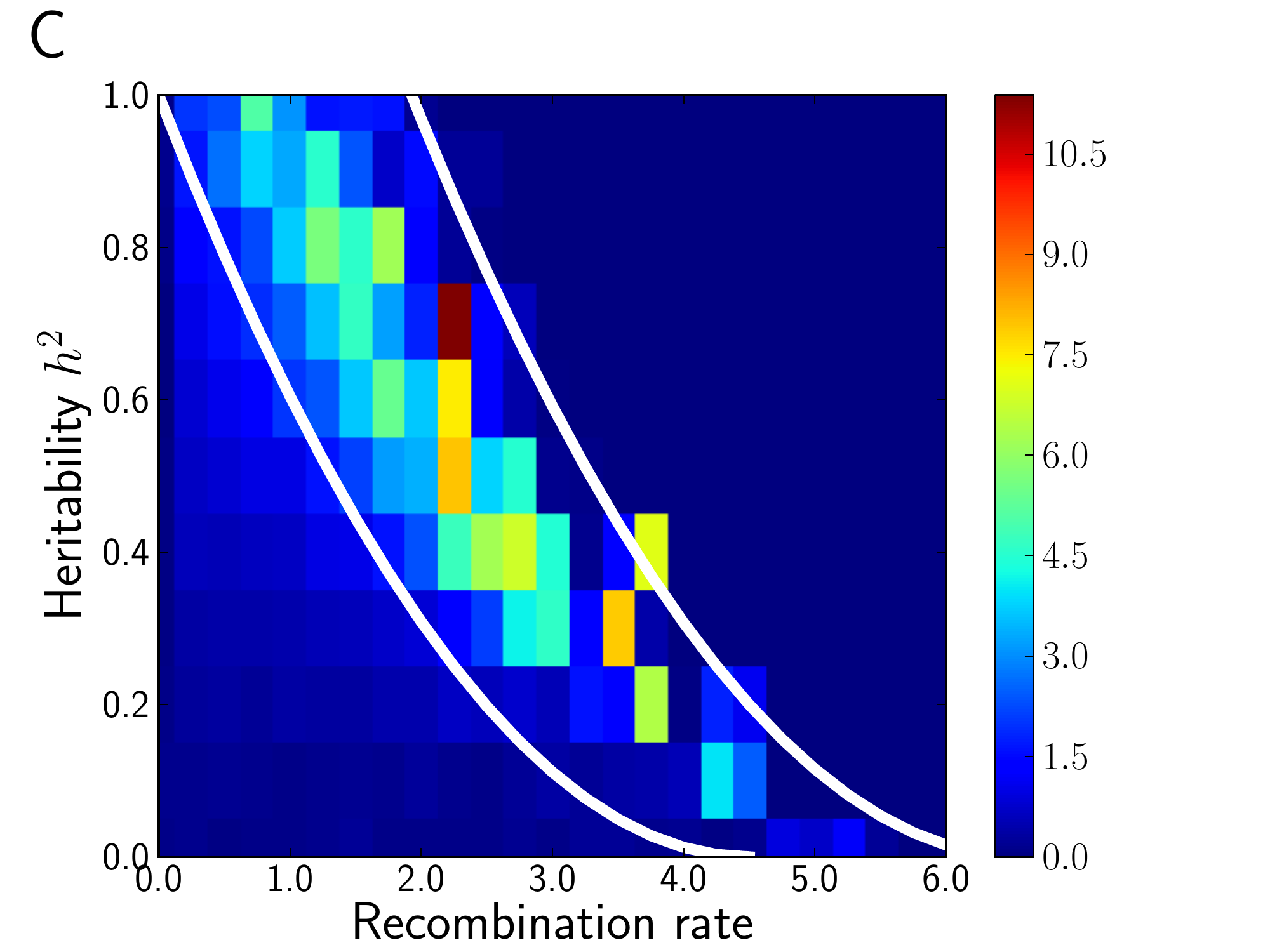}  
	\caption[labelInTOC]{Clonal condensation as a function of recombination rate and heritability.
	Panel A: Ratio of the traveling wave velocity to the total variance. At low heritability and low recombination rates, the average velocity of the traveling wave is much lower than the additive variance. The equality of velocity and additive variance promised by Fisher's Fundamental Theorem is recovered in the high recombination limit. Panel B shows the corresponding participation ratio $\log \langle Y_{\infty}(r,h) \rangle$. Low velocity goes along with high $\langle Y_{\infty}(r,h)  \rangle$. Panel C shows the coefficient of variation of the time trace $Y_t$, i.e., $\mathrm{std}(Y_t)/\mathrm{mean}(Y_t)$. $Y_t$ fluctuates strongly at intermediate recombination rates between the two critical lines identified in the main text and indicated by the white lines. In all panels, $N=10^{6}$ and $\sigma^2=0.0025$. The additive fitness is drawn from a Gaussian centered around the current mean additive fitness.}
  \label{fig:v_and_Y}
\end{center}
\end{figure}

\FIG{clones} shows that, at low recombination rate and heritability, the
population is always dominated by a few large clones with high non-heritable
fitness components, which produce a large number of (on average) non-fit
offspring. Rarely, such offspring are very fit, and replace their predecessors.
The probability that offspring are fitter than their parents, and with it
the velocity, increases dramatically with $h^2$ and $r$. Conversely, the size of
the fit clones and the participation fraction decreases with $h^2$ and $r$.
Simulation results for $Y$ and $v$ in the steady state are shown in
\FIG{v_and_Y} for different values of $h^2$ and $r$. In the following, we will
rationalize and quantify the observed behavior. To begin with, we will assume a
constant velocity and calculate $Y$. Again, these calculations are detailed in
Appendix \ref{sec:Ytraveling}.

After successful establishment, a clone will grow approximately
deterministically according to
\begin{equation}
\frac{\dot{n}_j}{n_j}  = F_j-r - v\tau_j - \mefit
\end{equation}
where $F_j=A_j+E_j$ is the sum of the additive fitness $A_j$, measured relative
to $vt$ when the clone was born, and $E_j$ is the epistatic fitness. The
advancing mean additive fitness is accounted for by $v\tau_j$, while $\mefit$ is the mean
epistatic fitness. The size of the clone peaks at age
$\tau_j^*=(F_j-\rE)/v$ (where we have defined $\rE = r+\mefit$) and then
decreases until the clone disappears. The maximal size of the clone is
$n_j^{max} \sim e^{-(F_j-\rE)^2/2v}$. Since the fittest genotypes in the
population have a fitness $\sqrt{2\log N}$ above the mean, they grow larger than
$N$ if
\begin{equation}
\rE < r^{*} = \sigma (1-\sqrt{\gamma})\sqrt{2\log N\sigma} \ ,
\end{equation}
which gives us a first indication of the breakdown of the mean field solution,
where $v=\sigma_A^2$. We confirm this again by calculating the integrals
$C_r(z,v)$ and $B_r(z,v)$ in Appendix \ref{sec:Ytraveling}; see
\EQ{app_Cr_traveling}.

Furthermore, we calculate $\langle Y(r,h^2)\rangle$ in  Appendix
\ref{sec:app_Yepi} and find that $\langle Y(r,h^2)\rangle$ starts to be larger
than $N^{-1}$ as soon as
\begin{equation}
\rE < r_c = \sigma(\sqrt{2}-\sqrt{\gamma})\sqrt{2\log N\sigma} \ ,
\end{equation}
These two conditions on $\rE$ define two critical recombination rates $r_c$ and
$r^{*}$ at which different features of the mean field solution break down. Note,
however, that this expression is not valid close to $v=0$, since we have assumed
a traveling wave with clones of finite lifetime.

\begin{figure}[htp]
\begin{center}
  \includegraphics[width=0.48\columnwidth]{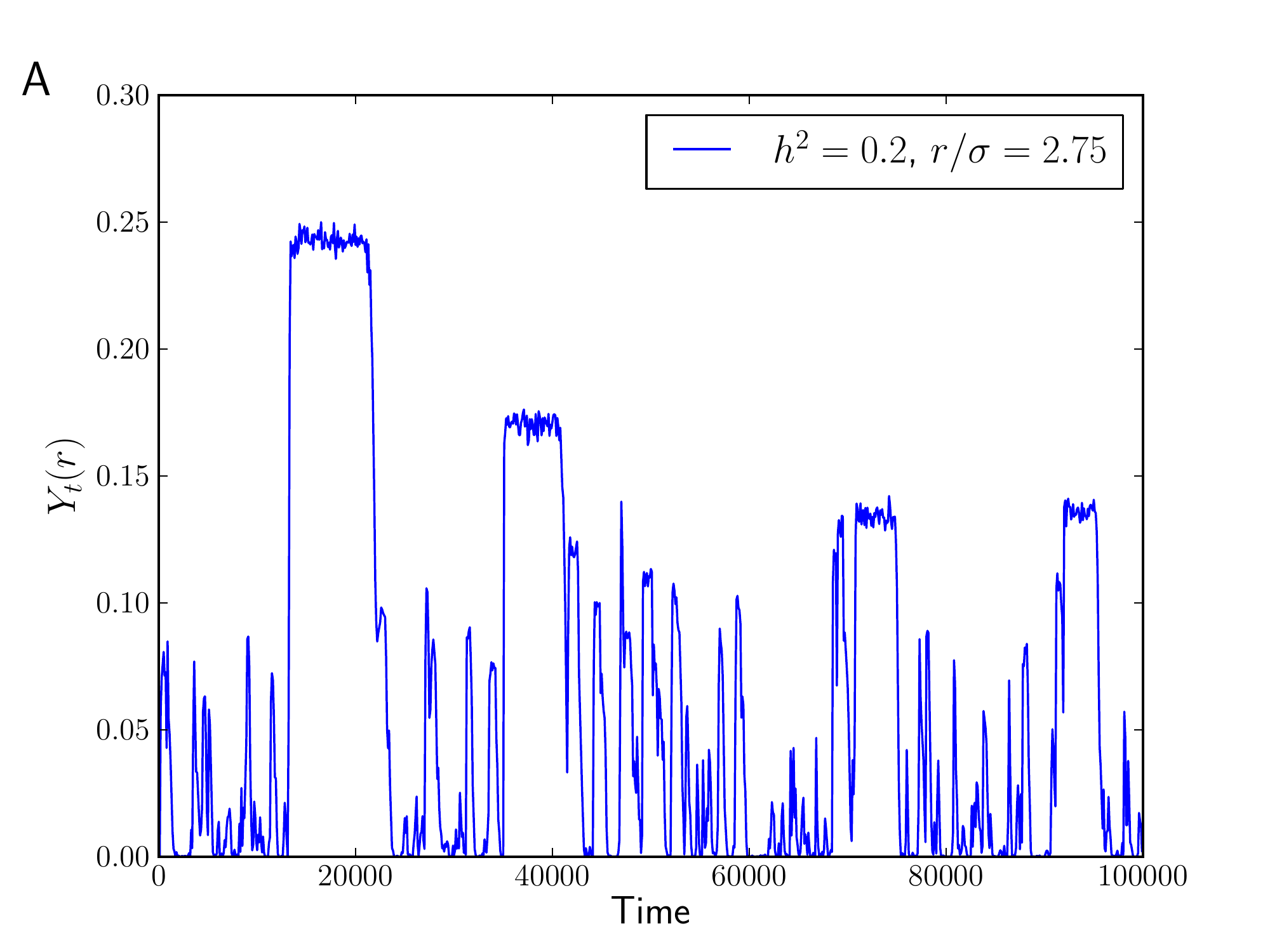}
  \includegraphics[width=0.48\columnwidth]{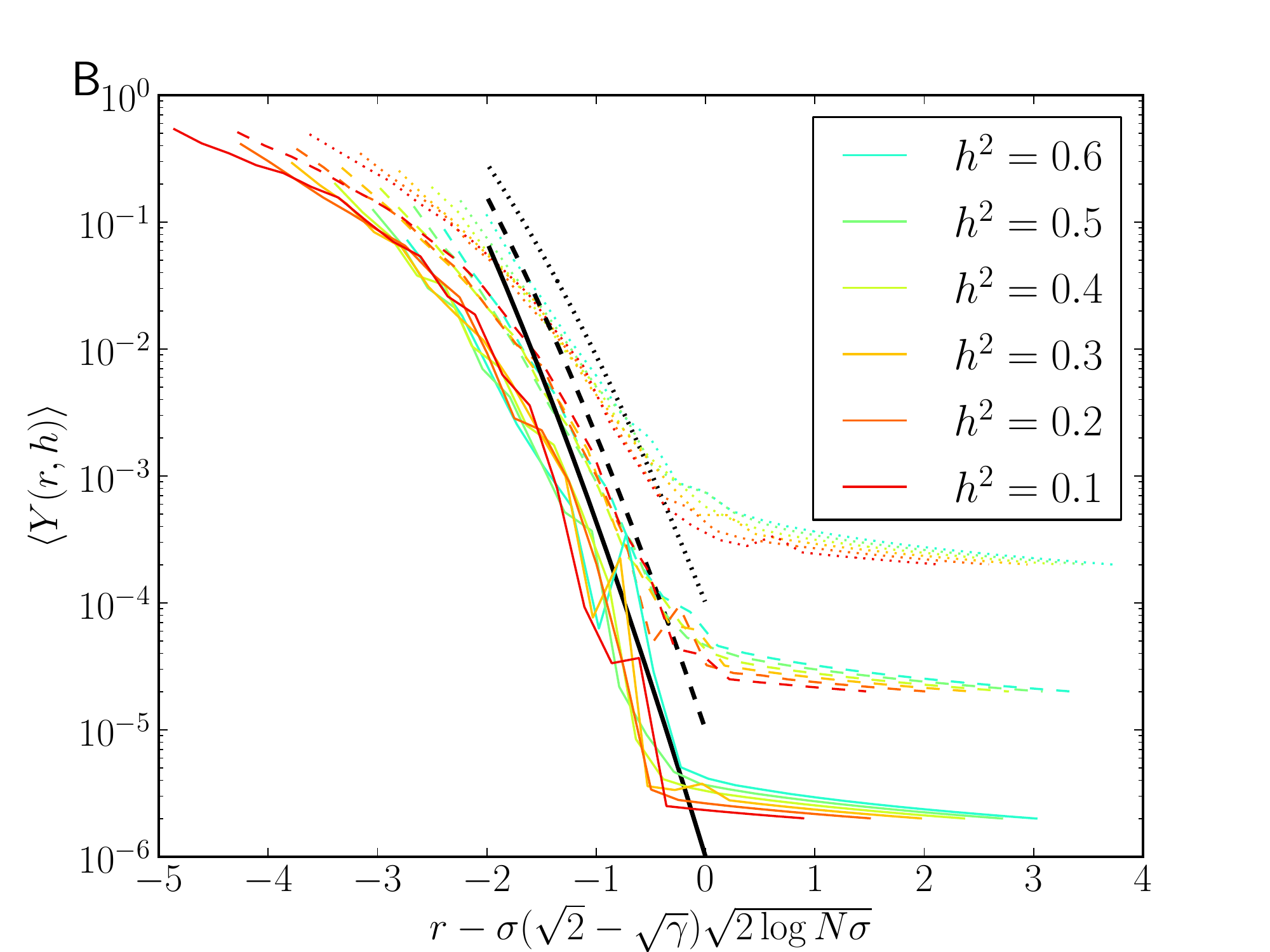}
  \caption[labelInTOC]{Panel A shows a time trace of $Y_t(r,h)$ which exhibits strong fluctuations. For part of the time, the population is dominated by a single clone and does not adapt. After a while, this clone is superseded and the population starts moving again. Panel B: Participation ratio for different heritabilities and different population sizes at an imposed velocity of $v=\sigma_A^2$ (the transition is sharper if $v$ is adjusted self-consistently). Note that above the critical recombination rate, $\langle Y\rangle$ is of order $N^{-1}$. The analytical predictions (\EQ{app_Yvlessone}), up to the prefactor which was fixed to ensure $Y(r_c, h^2)=N^{-1}$, are shown as black lines.  Dotted lines: $N=10^4$, dashed: $N=10^5$, solid: $N=10^6$, $\sigma^2=0.0025$.}
  \label{fig:stick_slip}
\end{center}
\end{figure}

Both of these two lines seem to play an important role for the behavior of the
population. Between the two lines, we observe a coexistence between condensed
populations and non-condensed populations, which results in large fluctuations of
$Y(t)$. An example trajectory of $Y(t)$ is shown in \FIG{stick_slip}. For a
limited amount of time, the population is condensed with large $Y$ and doesn't
move in the additive direction. It then becomes unstuck and adapts for a while
before getting stuck again.  The time average of $Y$ is dominated by these
intermittent condensates in this meta-stable region and is larger than $N^{-1}$
as soon as $r<r_c = \sigma(\sqrt{2}-\sqrt{\gamma})\sqrt{2\log N}$. Good simulation evidence for the different transition lines,
however, is difficult to obtain because of substantial sub-leading corrections.

To investigate the model in absence of this stick-slip behavior, we simulated a variant of
the model where additive fitness is drawn from a distribution that steadily moves towards
higher fitness with velocity $v$, as is assumed in the calculations. The time averaged 
$\langle Y(r,h)\rangle$ is shown in \FIG{stick_slip}B and compared to the predicted
onset of condensation by \EQ{app_Yvlessone} (black lines).

%

\section{Discussion}
\subsection{Summary}
Correlations between different parts of the genome are usually referred to as
linkage disequilibrium, suggesting that due to genetic linkage, i.e., a high
probability of coinheritance, the allele frequencies at different loci are not
independent. Here, we have used a different measure to characterize correlations
in the population. Instead of looking at correlations between individual loci, we
have characterized the distribution of clone sizes, or equivalently haplotype
frequencies, in adapting populations. Our analysis is not restricted to additive
fitness functions, but we have also analyzed a simple model where fitness is
partly heritable and partly random.

While macroscopic condensation, $\langle Y_{\infty} \rangle\sim \mathcal{O}(1)$,
sets in only for $r<\sigma/\sqrt{2\log N\sigma}$ in the absence of epistasis, we
observe condensation of the population at recombination rates of order $\sigma$
in presence of epistasis. Here, $\sigma$ is the standard deviation in fitness
and sets the strength of selection. The reason for this behavior is the fact
that the velocity at which the population adapts is smaller than the fitness
variance of the recombinant offspring. The additional epistatic variance allows
the seeding of new clones far ahead of the population mean, which is only slowly
catching up. Hence fit clones can out-grow any traveling Gaussian. At the same
time, condensed clones cause the average epistatic fitness to be significantly
greater than 0. Since this epistatic fitness is lost upon outcrossing, the
population has a tendency to partition into a few fit clones with high epistatic
fitness and a large number of recombinant genotypes with random epistatic
fitness. This co-existence between ``condensed'' and ``dust'' phases is seen in
\FIG{clones} in the panels on the right.
As long as the heritability, i.e., the fraction of additive variance, is larger
than zero, the population seeds new clones even at low recombination rates and
the rate of coalescence will be given by $\langle Y\rangle$ times the
characteristic turn-over rate of clones.

In the absence of any additive variance, the observed behavior is quite
different. In this case, the fitness function is completely random (a.k.a.
House-of-cards model \citep{Kingman:1978p31199}, or random epistasis/energy
model \citep{Derrida:1981p30865,Neher:2009p22302}). The only way the population
adapts is by sampling fitter and fitter individuals from the same distribution.
In other words, the population dynamics amounts to a record process where the
total number of samples taken increases as $N(1+rt)$. Records establish and grow
with the rate $E-\mefit-r$. One additional complication that is of particular
importance in the case $h^2=0$ is the fact that whenever a clone recombines with
itself, it does not generate a novel genotype. This has the tendency to shut off
recombination and stabilize clones as soon as they grow large, resulting in a
rapid loss of genetic diversity. In a previous
publication \citep{Neher:2009p22302}, some of us have studied the onset of
condensation in a more descriptive manner. Here, we have extended this work by
explicitly calculating $Y$, both during an initial transient as well as in a
steady state where variance is maintained.

The model we have used is extremely simplistic, and one might wonder about its
relation to real world populations. Nevertheless, it accounts for a number of
features of real populations such as heritabilities between $0$ and $1$,
outcrossing, and mimics a large number of loci in the sense of quantitative
genetics. These features give rise to qualitatively different dynamical regimes,
which will also be observable in more realistic models. Some facultatively
sexual populations are in fact remarkably close to this simple ``$E$ and $A$''
model. Many plants and microbial populations are facultative outcrossers. In
the event of outcrossing, a large number of crossovers on many chromosomes
produces many independently inherited genomic segments. HIV, for example,
recombines via template switching following coinfection at an outcrossing rate
of a few percent \citep{Neher:2010p32691,Batorsky:2011p40107}. In each of these
outcrossing events, roughly 10 crossovers are observed \citep{Levy:2004p23309}.
If populations are polymorphic at many loci, the resulting off-spring
distribution is approximately described by \EQ{rec_kernel}. We have made a
further simplification by assuming that the fitness of a recombinant offspring is
independent of its parents and simply drawn from a comoving distribution. This
assumption is expected to approximate recombination processes where the
offspring and parent fitness decorrelate rapidly over a few out-crossing events,
as for example in \EQ{rec_kernel} \citep{Neher:2010p30641}. Note, however, that
loci close together on the chromosome decorrelate only slowly.

The other dramatic simplification made above was the partitioning of fitness
into additive and random components corresponding to high order epistatic
interactions. More generally we expect to find epistatic interactions of various
orders, which are heritable to different extents. However, the ``$E$ and $A$''
model should not be thought of as a parameterization of the genotype-fitness
map, but rather as a partitioning into variance that can be explained by the
best fitting additive model, and the remaining variance
\citep{Fisher_1930,Neher:2011p45096}. The best-fitting additive model is in
general time dependent, and the heritability can change as the allele
frequencies change \citep{Turelli:2006p4414}. We know rather little about the
genetic architecture of fitness, which justifies the the use of such simple
models. In specific scenarios where the genotype-phenotype map is known, more
detailed modeling should be guided by the general conclusions drawn from the ``$E$
and $A$'' model.

The variation that we assume is always present among the recombinant off-spring,
it is ultimately fueled by mutations. The balance between the influx of
beneficial mutations and selection in facultatively sexual population has been
investigated in \citep{Cohen:2005p5007,Neher:2010p30641}. Even in sexual
populations, the dynamics of mutation can be strongly affected by the selection
through the clonal structure of the population.

\subsection{Why is $Y$ important?}
The participation ratio, $Y$, is exactly the probability for two individuals to
be genetically identical. Therefore, it immediately gives a measure of the clonal
structure of the population.  If the two genotypes are identical, they have had
a common ancestor in the recent past. Hence, $\langle Y\rangle$ is proportional
to the rate of coalescence, and as soon as $\langle Y\rangle$ is no longer
proportional to $N^{-1}$, coalescence is greatly accelerated relative to a
neutral model. It is well known that selection accelerates coalescence since fit
individuals have more offspring and dominate future generations. Recombination
tends to reduce the effects of linked selection since it decouples different
regions of the genome. We have calculated the rate of coalescence in our model,
which is set by a balance between selection and recombination. We have shown that
there is a critical recombination rate, where recombination is overwhelmed by
selection and the population structure changes qualitatively.

In the case of additive fitness functions, we have found that $\langle
Y\rangle\sim\exp\left[-\frac{r}{\sigma}\sqrt{2\log N\sigma}-\frac{r^2}{2\sigma^2}\right]$, which is in agreement with
earlier work \citep{Rouzine:2010p33121,Neher:2011p42539}. In this case, the
population consists of clones apparent as little ``bubbles'' in the
representation of \FIG{clones}. Any such bubble originates from a common
ancestor $\sim \sigma^{-1}\sqrt{2\log N\sigma}$ generations in the past, and $\langle
Y\rangle$ is the probability that two individuals belong to the same ``bubble''.
This bubble-coalescent is similar to ideas developed for structured populations
\citep{Nordborg:1997p43854} or the fitness class coalescent
\citep{Walczak:2011p45228}. Note, however, that the clone-size distribution is
very long tailed and the bubble coalescent is not in the universality class of
the Kingman coalescent \citep{Kingman:1982p28911}, but possibly of
Bolthausen-Sznitman type
\citep{Brunet:2007p18866,Neher:2011p42539,Mohle:2001p41279}.

Genetic identity between some of the individuals reduces the effect of
outcrossing, since identical parents produce identical offspring. Since the
probability of such an occurrence is equal to $Y$, the effective rate of
recombination in the partly clonal population is $r_{eff}=r(1-Y(r_{eff},h))$.
Hence, strictly speaking our calculations of $Y(r,h)$ deep in the clonal
condensation regime must be taken through a self-consistency step, which
replaces $r$ that was hereto an independent variable by a dependent variable
$r_{eff}$. This however would not change our estimates for $r_c(h)$ and
$t_c(r,h)$, which are defined by the point of first emergence of clones (when
$Y$ rises above $O(1/N)$).  Going beyond the MFT description of recombination,
one may define a participation ratio {\it{density}}
$\Upsilon(F)$ in terms of which $Y=\int dF \Upsilon(F)$ which picks up the
fitness dependence of $Y$:  individuals with relatively high fitness are much
more likely to be clonally related.

The significance of $Y$ is not limited to the case of exact genetic identity
within clones. In particular, mutations would introduce additional polymorphic
loci into the ``clones" that were the focus of our study. Yet the genetic
structure of the population introduced by clonal condensation still survives:
one only needs to distinguish between high-frequency polymorphisms, which are
being reshuffled by recombination as approximated by our model and the low
frequency new polymorphisms due to recent (on the time scale of clonal growth)
mutations. The latter would appear on the background that is clonal with respect
to the high-frequency alleles. Thus the ``clones" emerging at low $r$ should be
thought of as haplotypes \citep{McVean:2005p46450} and the ``clonal
condensation" is the process that suppresses the number of haplotypes on small
length scales.

The participation ratio can be readily generalized to allow for a degree of genetic
distance within a pair of individuals.
As currently defined $Y=\langle \delta (||g-g'||) \rangle$ (where $||g-g'||$
stands for the genetic distance between $g$ and $g'$). This is immediately
recognizable as a special case of the Parisi order parameter $q(x)=\langle
\delta (||g-g'|| -x) \rangle$ \citep{Mezard:2009p28797}. The latter therefore
provides an interesting representation of the haplotypic structure of
populations. It would be interesting to understand whether more realistic models
of fitness landscapes (with low order, rather than random epistasis) would
generate more complex hierarchical structure of $q(x)$ than the simple
``dust/clone" dichotomy found in our simply model. The relation between the REM
and Sherrington-Kirkpatrick models of spin glasses
\citep{Sherrington:1975p36835} could provide useful guidance and ultimately
yield better understanding of haplotype distributions and recombinant coalescents
\citep{Hudson:1995p18197,Stephens:2001p46309,Myers:2003p46466}.

\subsection{Future Directions}
The analysis of ``clonal condensation" presented above can and should be
extended in several ways.
Within the confines of the model considered, one may want to obtain
a better understanding of the ``mixed phase" lying between the two transition
lines identified in  \FIG{v_and_Y}. This phase is characterized by large
fluctuations in clone size distribution, and hence in $Y$, even in the
approximation imposing a fixed traveling wave velocity $v$. In reality, the
population sets its own instantaneous rate of change of average fitness, which
depends sensitively on the fitness of the leading clones and changes with time
as new fitness ``records" are established by fresh recombinants. We have already
described in \FIG{stick_slip} the ``stick-slip" dynamics, which is characteristic
of the mixed phase regime. (Needless to say, the existence of the mixed phase
region corresponds to the 1st order nature of the clonal condensation transition
for $h \neq 0,1$.  )  A fully quantitative description of this behavior would
require going beyond MFT. So far, attempts to describe fluctuations in the dynamics of adaptation 
have been few and far apart
\citep{Brunet:2007p18866,Hallatschek:2011p39697,Neher:2012p47353}:
a quantitative description of the ``stick-slip" progress of adaptation would
represent a major step forward.

Another necessary extension of the model involves mutational influx. A non-zero
mutation rate would provide an influx of genetic variation and define true
statistically stationary states corresponding to adaptive traveling waves
\citep{Tsimring:1996p19688,Rouzine:2003p33590,Desai:2007p954,Cohen:2005p45154,Neher:2010p30641}
or, in the presence of both deleterious and beneficial mutations, to a dynamic
mutation selection balance \citep{Goyal:2011p45049}. In that case, emergent
clones become ``fuzzy" as they accumulate mutations,
and the participation ratio should be replaced by a more general Parisi order
parameter as explained above. The result should provide interesting quantitative
insight into the expected genetic structure of facultatively sexual populations.

Perhaps the most interesting and important extension of the present work would
involve a generalization of the model and its analysis to linear genomes and
obligatory sexual populations. In contrast to the current model where
recombination freely re-assorts the loci, a more realistic linear chromosome
model  would describe recombination in terms of relatively infrequent crossover.
This would naturally tie the frequency of recombination to the length of the
segment considered. We expect that on sufficiently short scales, where
recombination is infrequent, a tendency for haplotype condensation would be
manifest if the population is diverse enough. Whether epistasis plays a
significant role in the condensation process will depend on the distribution of
epistatic interactions along the genome \citep{Neher:2011p45096}. If there is a
strong tendency of mutations to interact with mutations nearby
\citep{Callahan_2010}, the heritability decreases as smaller and smaller
segments are considered. This could result in condensation of mutations into
``super-alleles''. However, the embedding of the haplotype in question into a larger 
genome gives rise to complications related to Hill-Robertson interference
\citep{Hill:1966p21029,Hudson:1995p18197,Barton:1994p34628}.
Transient associations with other genomic regions will either boost
(the hitch-hiking process) or suppress (background selection) the spread of a
haplotype in the population. This reduces the efficacy of selection and 
gives rise to a stochastic dynamics with rather different properties 
than genetic drift \citep{Neher:2011p42539}. Bridging the different scales and resulting
dynamical regimes represent an important challenge for the future.

\subsection{Conclusion.}
In conclusion, we stress the distinction between the QLE and ``clonal
condensation" regimes. In the QLE regime, recombination is sufficiently rapid to
overwhelm any clonal amplification due to selection, and correlations between
alleles at different loci are relatively weak. In this regime, the correlations
between loci are well described by linkage disequilibrium, which measures
population averaged pair correlation of loci. By contrast, clonal condensation
is a non-perturbative, large deviation from linkage equilibrium (under which
loci are completely uncorrelated), which in particular results in a
stratification of the population depending on its fitness: clones appear
predominantly in the upper reaches of the fitness distribution. Strong
heterogeneity among individuals along the fitness axis is not captured by
measuring linkage disequilibrium and other traditional approaches. Understanding
strong interactions in multi-locus systems requires new ideas and tools.  We
have found simple models such as the REM to be a very useful source of insight
into these non-trivial aspects of population genetics.
\acknowledgements We are grateful for stimulating discussions with H.~Teotonio,
A.~Dayarian, I.~Rouzine, D.~Fisher, M.~Desai and M.~Lynch and would like to thank
T.~Kessinger for careful reading of the manuscript.  RAN is supported by
an ERC-starting grant 260686 and BIS acknowledges support of the HFSP
RFG0045/2010.

\appendix
\section{Participation ratio for Clonal Condensation.}
\label{sec:app_definitions}
Following its definition for the REM we define the participation ratio for a set of clones as the sum of squared frequencies 
\begin{eqnarray}
\label{eq:app_YREM}
\langle Y_t\rangle &=&\langle \sum_i \left(\frac{n_i(t)}{\sum_j n_j(t)}\right)^{2}\rangle_{\{F_k\}}= 
\sum_i \int_0^\infty dz\; z \ \langle \ n_i^2(t) e^{-z n_i(t)} \prod_{j \neq i} e^{-z n_j(t)} \ \rangle \ ,
\end{eqnarray}
where
\begin{eqnarray}
\label{eq:app_clone_growth_general}
n_{F_i, \tau_i}(t)=e^{(F_i -r)\tau_i-\int_{t-\tau_i}^{t} d\tau' \mfit_{\tau'} } 
\end{eqnarray}
describes the size of clone $i$  created a time $\tau_i$ in the past with fitness $F_i$. Due to stochastic effects, most clones go extinct early on, and the fraction $\sigma$ that remains is on average larger by a factor of $\sigma^{-1}$. We will ignore this correction and reinstantiate it later when comparing to simulations. The rate of growth of the clone is controlled by the ``replication rate": the fitness  relative to the  time dependent average fitness of the population $\mfit(t)$ minus the recombination rate. The average over clones $\langle \ ... \ \rangle$ implies averaging over $F_i$ and $\tau_i$. It can be decomposed into the average over the initial population (of individuals present at $t=0$ with $\tau_i=t$ ) and the average over subsequent recombinants that are generated via Poisson process with rate $Nr$. Hence
\begin{eqnarray}
\langle \ \prod_{j} e^{-z n(t,F_j, \tau_j)} \rangle 
&=&\left \{ \int dF \rho(F) e^{  -z n_{F, t}(t)} \right \}^N \times 
\sum_{k=0}^{\infty} { e^{-Nrt} \over k!}  \left \{ Nr \int_0^t d\tau \int dF \rho(F) 
e^{ -z n_{F, \tau}(t)} \right \}^k\\
&=&\left \{ 1-\int dF \rho(F) \left (1- e^{  -zn_{F, t}(t)} \right ) \right \}^N\times
\exp \left[- Nr \int_0^t d\tau \int dF \rho(F) \left(1-e^{ -zn_{F, \tau}(t)} \right) \right] \\
\\
&\approx &\exp\left[{-NC_0 (z,t)-NC_r (z,t)}\right] \ ,
\end{eqnarray}
where we have defined
\begin{eqnarray}
C_0(z,t)&=& \int dF \rho(F) \left [1- e^{  -zn_{F, t}(t) } \right ]   \ ,
\end{eqnarray}
which expresses the average over the initial population and
\begin{eqnarray}
\label{eq:def_Cr}
C_r (z,t)&=& r   \int_0^t d\tau \int dF \rho(F)  \left [1-e^{ -zn_{F, \tau}(t)} \right ] 
\end{eqnarray}
which gives the contribution of all recombinants.
If we further define
\begin{eqnarray}
B_0(z,t)&=& \int dF \rho(F) n_{F, t}^2(t) e^{-zn_{F, t}(t)} = - \frac{\partial^2}{\partial z^2}C_0(z,t) \ ,
\end{eqnarray}
\begin{eqnarray}
\label{eq:def_Br}
B_r(z,t)&=&r \int_0^t d\tau \int dF \rho(F) n_{F, \tau}^2(t)e^{-zn_{F, \tau}(t)}=- \frac{\partial^2}{\partial z^2}C_r(z,t) \ ,
\end{eqnarray}
we arrive at the following expression for the participation fraction:
\begin{eqnarray}
\langle Y_t\rangle &=&N\int_0^{\infty} dz \ z  [B_0(z,t)+B_r(z,t)]e^{- N(C_0(z,t)+C_r(z,t))}
\end{eqnarray}
At sufficiently long times, $rt \gg 1$, the population is dominated by new recombinants so that $B_0$ and $C_0$ may be neglected in comparison with $B_r$ and $C_r$. 

The growth and decay of clones, \EQ{app_clone_growth_general}, depends critically on the dynamics of the mean fitness, which in turn depends on the heritability of fitness. We have discussed the different cases ($h^2=0,1$ and intermediate heritabilites) in the main text and will present the calculations pertaining to them in separate appendices below. To streamline the calculation, we will rescale all
rates with $\sigma$ ($r\to r/\sigma$, $A\to A/\sigma$, $E\to E/\sigma$, $F\to F/\sigma$) and multiply times with $\sigma$ ($t\to t\sigma$, $\tau \to \tau\sigma$). Consequently, $v$ now equals $\gamma$. We will reinstantiate $\sigma$ in the final results to facilitate comparison with formulae in the main text. In addition, the following integrals will prove useful:
\begin{eqnarray}
\int dx e^{a x + 2b x-e^{b x}} &=& \int_0^\infty \frac{d\xi}{b} \xi^{1-\frac{a}{b}}e^{-\xi} = \frac{1}{b}\Gamma(2-\frac{a}{b}) \\
\int dx e^{a x}(1-e^{-e^{b x}}) &=& \int_0^\infty \frac{d\xi}{b} \xi^{-1-\frac{a}{b}}(1-e^{-\xi}) = \left[-\frac{1}{a}\xi^{-\frac{a}{b}}(1-e^{-\xi})\right] +  \int_0^\infty \frac{d\xi}{a} \xi^{-\frac{a}{b}}e^{-\xi} \\ \nonumber
&=& \frac{1}{a}\Gamma(1-\frac{a}{b}) = -\frac{1}{b}\Gamma(-\frac{a}{b})
\end{eqnarray}
where the latter holds only for $a<b$.

\section{Participation ratio for the fully epistatic $h=0$ case}
\label{sec:app_hzero}
The fully epistatic case with recombination or mutation is the closest relative of the REM  in population genetics, known in the population genetics literature as the ``house-of-cards" model \citep{Kingman:1978p31199}. In this model, every new genotype is drawn from the same distribution which we will take to be the standard normal distribution, that is, fitness is non-heritable.

Let us consider $rt \gg 1$ so that enough recombinants are produced to dominate over the initial set of genotypes and evaluate $C_r(z,t)$ and $B_r(z,t)$. Furthermore, let us assume that the mean fitness is constant (we will revisit this assumption later). The size of a clone with age $\tau$ is then simply $n=\exp[(E-\rE)\tau]$, and $C_r(z,t)$ is given by
\begin{eqnarray}
C_{r,t}(z)&\approx&r\int_0^t d\tau \int_{-\infty}^{\infty} {dE  \over \sqrt{2 \pi}} e^{-{E^2 \over 2}}   [1-e^{-ze^{(E-\rE)\tau }}] \ .
\end{eqnarray}
Our strategy in approximating this integral will be to identify the region where $\Phi=ze^{(E-\rE)\tau }<1$ and expand the $e^{-\Phi}$ exponential in that region; outside that region we shall neglect the exponential compared to $1$ (in the square bracket). Defining $\theta = E-\rE$, we have
\begin{eqnarray}
C_{r,t}(z)&\approx&r \int_{-\infty}^{\infty} {d\theta  \over \theta \sqrt{2 \pi}} e^{-{(\theta+\rE)^2 \over 2}}  \int_0^{t\theta} d\tau [1-e^{-ze^{\tau }}]\\
&\approx& zr \int_{-\infty}^{0} {d\theta  \over \theta \sqrt{2 \pi}} e^{-{(\theta+\rE)^2 \over 2}}  [e^{\theta t }-1]+zr \int_0^{t^{-1}\log z^{-1} } {d\theta  \over \theta \sqrt{2 \pi}} e^{-{(\theta+\rE)^2 \over 2}}  \int_0^{t\theta} d\tau e^{\tau }\\
&+&r \int_{{t^{-1}\log z^{-1} }}^{\infty} {d\theta  \over \theta \sqrt{2 \pi}} e^{-{(\theta+\rE)^2 \over 2}}  \int_0^{t\theta} d\tau [1-e^{-ze^{\tau }}]\\
&\approx&z  [1-e^{-\rE t }]+zr \int_0^{t^{-1}\log z^{-1} } {d\theta  \over \theta \sqrt{2 \pi}} e^{-{(\theta+r)^2 \over 2}}[e^{t\theta} -1] +{r  e^{-{E_*^2 \over 2}}  \over (E_*-r) E_* \sqrt{2 \pi}}   [(1-z)+{t \over E_*}]\\
&\approx&z  +{rt  e^{-{E_*^2 \over 2}}  \over (E_*-\rE) E_*^2 \sqrt{2 \pi}}   \ ,
\end{eqnarray}
where $E_*=t^{-1}\log z^{-1}+\rE$ and we have along the way assumed that  $t>E_*(t)$, which is realized for $t>t_*={\rE \over 2} [1+\sqrt{1+{4\over \rE^2}\log z^{-1}}]$. As we shall see below, the dominant contribution to $Y$ comes from the region $z^{-1} \sim N$, so that in the low $r$ regime ($\rE<\sqrt{2 \log N}$) we have $t_* \approx \sqrt{\log N}$. At the end of the calculation we shall check that $t>t_*$ condition is satisfied. 

At short times the second term in our expression for $C_{r,t}(z)$ is clearly smaller than the first, so that the $dz$ integration in the expression for $Y$ is controlled by the $e^{-Nz}$ factor, which set the scale for $z$ at $z \sim N^{-1}$.
Let us next estimate the time for which the second term in the expression for $C_{r,t}(z)$ becomes larger than the first provided that $z \approx N^{-1}$. We obtain
\begin{eqnarray}
e^{{(t^{-1}\log N +\rE )^2 \over 2}}  <  {rt^2N   \over   (t^{-1}\log N +\rE )^2 \log N \sqrt{2 \pi}}  \ .
\end{eqnarray}
Let us define $\hat{t}=t/\sqrt{{ 1\over 2} \log N}$ and  $\hat{r} =r/\sqrt{2 \log N}$; then
\begin{eqnarray}
N^{-1+(\hat{t}^{-1} +\hat{r })^2}  <  {\hat{r }\hat{t }^2   \over   \sqrt{2 \log N}(\hat{t}^{-1} +\hat{r })^2 \sqrt{2 \pi}}    \ .
\end{eqnarray}

For $\hat{r}>\hat{r}_c=1$ this can only be satisfied if $\hat{t}^2 >\hat{t}_*^2 \sim N^{\hat{r}^2-1}\sqrt{ \log N}$, which diverges in the large $N$ limit.
Below $r_c$, but still close to it, we can approximate the crossover $ \hat{t}_* \gg \hat{r }^{-1}$, and the  crossover condition becomes
\begin{eqnarray}
\hat{t}_*^{-1}  \approx  {1-\hat{r }^2\over 2\hat{r }}+{\log \left [ {\hat{t }_c^2   \over  4 \hat{r }\sqrt{\pi \log N} }   \right ] \over 2\hat{r }\log N} \ ,
\end{eqnarray}
which is an approximation valid for $r \sim r_c=\sqrt{2 \log N}$.

It remains to calculate the participation ratio as a function of $t$ asymptotically for $t \gg t_c$ and for $t \approx t_c$. To do so, we need to evaluate the $B$-integral and perform $dz$ integration in the integral representation for $Y(t)$. The $B$-integral can be obtained by differentiating $C_r(z,v)$ twice with respect to $z$, which evaluates approximately to 
\begin{eqnarray}
&B(z,t)=-z^{-2}\left [ t^{-2} {d^2 \over dE_*^2}+t^{-1} {d \over dE_*} \right ] 
\left \{ {rt  e^{-{E_*^2 \over 2}}  \over (E_*-r) E_*^2 \sqrt{2 \pi}}   \right \} \approx z^{-2} {r  e^{-{E_*^2 \over 2}}  \over (E_*-r) E_* \sqrt{2 \pi}}  \ .
\end{eqnarray}
Hence
\begin{equation}
\begin{split}
\langle Y \rangle &=  N\int_{0}^{\infty} \frac{dz}{z} \frac{re^{-{{E_*}^2 \over 2}}} {(E_*-r) E_*\sqrt{2\pi}} e^{-Nz- N{rt  e^{-{E_*^2 \over 2}}  \over (E_*-r) E_*^2 \sqrt{2 \pi}}} \\
&\approx  \int_{0}^{\infty} {du}  e^{-Ne^{-t(u-r)}}  {d \over du}e^{- N{rt  e^{-{u^2 \over 2}}  \over (u-r) u^2 \sqrt{2 \pi}}}\\
&\approx  \int_{0}^{\infty} {d\phi}  e^{-Ne^{-t(u(\phi)-r)}-\phi}  \approx e^{-Ne^{-t(u(1)-r)}} 
\end{split}
\end{equation} \ ,
where
\begin{eqnarray}
\phi={Nrt  e^{-{u^2 \over 2}}  \over (u-r) u^2 \sqrt{2 \pi}}
\end{eqnarray}
and 
\begin{eqnarray}
u(\phi)=\sqrt{ 2 \log \left [ {Nrt   \over  \phi (u-r) u^2 \sqrt{2 \pi} } \right ] }
\end{eqnarray}
so that
\begin{eqnarray}
u(1) \approx \sqrt{ 2 \log N} \sqrt{ \left [ 1+ { \log crt \over \log N}  \right ] }
\end{eqnarray}
which upon substitution into Eq. B12 yields the final result
\begin{equation}
\begin{split}
\langle Y \rangle & \approx \exp\left[-N\sigma e^{-\sqrt{2\log N\sigma}  (1-\frac{r}{r_c})t\sigma }\right] \ .
\end{split}
\end{equation}
 This result deviates from zero for $t>t_c(r) = \frac{r_c\sqrt{2\log N\sigma}}{2\sigma(r_c-r)}$ consistent with our expected $t_c$ for $r \approx r_c$. For $r>r_c$ the participation ratio stays small for all $t$.

\section{Participation ratio for the general traveling wave}
\label{sec:Ytraveling}
As before, the participation fraction is given by an integral over clones seeded in the past. As opposed to the fully epistatic case, however, we now consider a finite velocity, which limits the lifetime of clones. A clone seeded at time $\tau$ in the past at fitness $A$ above the mean has a size
\begin{equation}
n_j = e^{(F_j-\rE)\tau_j - \frac{v\tau_j^2}{2}} \ ,
\end{equation}
where $\rE=r+\mefit$.
New clones get seeded with rate $Nr$, and to calculate $\langle Y \rangle$, we need to evaluate the integrals $C_r(z,v)$ and $B_r(z,v)$ as defined in \EQS{def_Cr}{def_Br}. In contrast to the case discussed above with $v=0$, a finite $v$ causes clones to have a limited life-time, even if they are initially very fit. 

The integral $C_r(z,v)$ over $F$ and $\tau$ has one contribution from young (small) clones with average fitness, in which case $z n_j$ is small. This young contribution evaluates to
\begin{equation}
C_r^y(z,v) = z r \int_0^\infty d\tau \int \frac{dF}{\sqrt{2\pi}} e^{-\frac{F^2}{2} + (F-\rE)\tau - \frac{v\tau^2}{2}} \approx \frac{rz}{\rE} \ .
\end{equation} 
Here we have evaluated only the contribution from young clones and neglected the fact that there is a growing quadratic term in the exponent. The latter is due to old clones; we will evaluate this term next. If $v<1$, the integrand starts growing again with $\tau$ and $\theta$ until it is cutoff at $\log z^{-1}  = \theta\tau - \frac{v\tau^2}{2}$. For a given $\tau$, this cut-off is at $\theta^{*} = \tau^{-1}\log z^{-1} + v\tau/2$, and the integrand has a peak at $\theta^*$ if $(1-v)\tau > \rE$. In this case, we expand the integrand around $\theta^*$, i.e., $F^{*}=\theta^{*}+\rE$, and find for the old clones
\begin{equation}
\begin{split}
C_r^o(z,v)&\approx  r\int_0^\infty \frac{d\tau}{\sqrt{2\pi}} e^{-\frac{{F^{*}}^2}{2}}\int \frac{d\delta \theta}{\sqrt{2\pi}} e^{F^{*}\delta \theta}(1-e^{e^{\tau\delta \theta}}) = -r\int_0^\infty \frac{d\tau}{\sqrt{2\pi}\tau} e^{-\frac{{F^{*}}^2}{2}}\Gamma(-\frac{F^*}{\tau}) \ ,
\end{split}
\end{equation}
The remainder can be evaluated by assuming that the non-exponential parts are slowly varying. The integrand peaks roughly at $\frac{\log z^{-1}}{{\tau^{*}}^2}=\frac{v}{2}$ or $\tau^{*} = \sqrt{2v^{-1}\log z^{-1}}$. This translates into $F^* = \sqrt{2v\log z^{-1}}+\rE$,  and the curvature is $2F^*\log z^{-1}/{\tau^{*}}^3=vF^*/\tau^{*}\approx v^2$. Hence  
\begin{equation}
\begin{split}
\label{eq:app_Cr_traveling}
C_r^o(z,v)&\approx  -\frac{rz^v e^{-\rE\sqrt{2v\log z^{-1}}-\frac{\rE^2}{2}}}{\sqrt{2v\log z^{-1}}}\Gamma\left(-v(1+\frac{\rE}{\sqrt{2v\log z^{-1}}})\right) \ .
\end{split}
\end{equation}

$B_r(z,v)$ can be evaluated by differentiating $C$ twice, which yields
\begin{equation}
\label{eq:app_Br_traveling}
B_r^o(z,v)\approx \frac{rz^{v-2}e^{-\rE\sqrt{2v\log z^{-1}}-\frac{\rE^2}{2}}}{\sqrt{2v\log z^{-1}}}\Gamma\left(2-v(1+\frac{\rE}{\sqrt{2v\log z^{-1}}})\right) \ .
\end{equation}

\subsection*{Additive fitness functions}
\label{sec:app_Yadd}
In the additive case with $v=1$,  the cut-off $C_r(z,v)$ is dominated by the young clones for any recombination rate. Its second derivative, however, is dominated by the contribution from old clones if $r<(\sqrt{2}-1)\sqrt{2\log N}$. In this regime, we find
\begin{equation}
\langle Y(r,1)\rangle \approx N \int_0^\infty dz \frac{r e^{-r\sqrt{2\log z^{-1}}-\frac{r^2}{2}}}{\sqrt{2\log z^{-1}}}\Gamma(1-\frac{r}{\sqrt{2\log z^{-1}}}) e^{-Nz} 
\sim  e^{-\frac{r}{\sigma}\sqrt{2\log N\sigma}-\frac{r^2}{2\sigma^2}} \ .
\end{equation}
In the last step, we have dropped all preexponential factors and reinstantiated $\sigma$. This correction approximately accounts for the fact that only a fraction $\sigma$ of the clones that are seeded are successful.

For larger $r$, even the $B_r(z,h)$ term is dominated by young clones and,  $Y\sim N^{-1}$. This behavior holds while the recombination rate is larger than $1/\sqrt{2\log N}$. For smaller $r$, the velocity starts to deviate from $1$. In this case, $C_r(z,h)$ is dominated by old clones. For large enough populations, the powers of $z$ vary much more quickly than all other terms, which we denote collectively by $A(z)$, and we can approximate the integral as 
\begin{equation}
\begin{split}
\label{eq:app_Yvlessone}
\langle Y(r,h)\rangle & = -\int_0^\infty dz z^{v-1}A(z)\Gamma(2-\frac{v(\theta^*+r)}{\theta^*})  e^{z^v A(z)\Gamma(-\frac{v(\theta^*+r)}{\theta^*})} \\ 
& = -v^{-1} \frac{\Gamma(2-\frac{v(\theta^*+r)}{\theta^*})}{\Gamma(-\frac{v(\theta^*+r)}{\theta^*})} = 
 \frac{\theta^*+r}{\theta^*}(1-\frac{v(\theta^*+r)}{\theta^*}) \approx 1-v = 1-\gamma \ ,
\end{split}
\end{equation}
where in the last step we reintroduced $\sigma$ by replacing $v$ with $\gamma = v/\sigma^2$. Along the way, we have assumed $r\ll \sigma\sqrt{2\log N\sigma}$, which is consistent with the assumption of small $r$ made above.

\subsection*{Epistatic fitness function}
\label{sec:app_Yepi}
In cases with heritabilities $0<h^2<1$, we generally have $v<1$ and can be in a regime where $C_r(z,v)\approx z$, while $B_r(z,v)$ is dominated by old clones. In this case, we have
\begin{equation}
\langle Y(r,h)\rangle \approx N \int_0^\infty dz z^{v-1}\frac{r e^{-r\sqrt{2v\log z^{-1}}-\frac{r^2}{2}}}{\sqrt{2v\log z^{-1}}} e^{-Nz} \sim  N^{1-\gamma} e^{-\frac{r}{\sigma}\sqrt{2v\log N\sigma}-\frac{r^2}{2\sigma^2}} \ .
\end{equation}
This starts to be of order $N^{-1}$ as soon as $\rE<\sigma(\sqrt{2}-\gamma)\sqrt{2\log N\sigma}$. This is similar to the condition identified in \citep{Neher:2009p22302} based on the factorization of the mean field solution. Note, however, that this expression does not hold near the $v=0$ line, since it is derived assuming a steady traveling wave.

\section{Velocity in the low recombination limit}
\label{sec:rouzine}
With sufficiently frequent recombination, the mean fitness in the population increases with a rate $v$ given by the variance in additive fitness among recombinant offspring. At lower recombination rates, however, the fitness variance in the population gets reduced below that of the distribution of recombinants. With reduced the additive variance, $v$ decreases. This effect has been studied in \citep{Rouzine:2005p17398} for a model with only additive fitness, and we reproduce this argument for our simplified recombination model. 

We will again work in units where the variance of the fitness distribution of recombinant genotypes is 1 and measure fitness $a$ relative to the instantaneous mean fitness in the population. The invariant fitness distribution $P(a)$ in the comoving frame $a = A-vt$ is governed by 
\begin{equation}
-v\partial_a P(a) = (a-r)P(a) + \frac{r
e^{-\frac{a^2}{2}}}{\sqrt{2\pi}}
\end{equation}
where the last term accounts for the injection of recombinant offspring. 
This equation has the solution
\begin{equation}
\label{eq:Padditive}
P(a)=rv^{-1} e^{-\frac{(a-r)^2}{2 v}}
\int_{a}^{a_c} \frac{da'}{\sqrt{2\pi
}}
e^{-\frac{a'^2}{2}+\frac{(a'-r)^2}{2 v}} \ .
\end{equation}
Note that this solution becomes negative for $a>a_c$, and only the part
$a<a_c$ is of interest. The zero crossing $a_c$ marks the
position of the most fit individuals in the population, and its value will depend
on the population size. 

To determine the velocity of a finite
population, \citet{Rouzine:2005p17398} compared the rate at which new
genotypes ahead of $a_c$ (records) are created to the
speed at which the bulk of the population moves towards higher fitness.

Within the model, recombinant genotypes are generated with rate $rN$ and drawn
from a Gaussian distribution with unit variance. Hence genotypes with
$a>a_c$ are produced at rate
\begin{equation}
rN\int_{a_c}^{\infty}
\frac{da}{\sqrt{2\pi}}e^{-\frac{a^2}{2}} \approx
\frac{rN}{a_c\sqrt{2\pi}}e^{-\frac{a_c^2}{2}} \ .
\end{equation}
Such a genotype seeded $a_c$ ahead of the mean has a chance $a_c \sigma$ to establish and, if established, advances the nose by $\frac{1}{a_c}$. We therefore find for the speed of the nose
\begin{equation}
\label{eq:vnose}
v_{nose} \approx \frac{\sigma rN}{a_c}e^{-\frac{a_c^2}{2}}
\end{equation}
To solve for $v_{nose}=v$ and $a_c$, we need an
additional relation which can be obtained from the condition that $P(a)$ is
normalized. In the limit of small $r$, we can approximate \EQ{Padditive} 
\begin{equation}
P(a) = \frac{r}{v}e^{-\frac{(a-r)^2}{2 v}}
\int_{a}^{a_c} \frac{da'}{\sqrt{2\pi}}
e^{\frac{(1-v)a'^2}{2v}-\frac{ra'}{v}+\frac{r^2}{2
v}} \approx
\frac{e^{-\frac{(a-r)^2}{2v}}}{\sqrt{2\pi
}}\frac{r}{(1-v)a_c-r} e^{\frac{(1-v)a_c^2}{2v}-\frac{ra_c}{v}+\frac{r^2}{2v}}
\end{equation}
The approximation is valid for $a_c (1-v)\gg r$; otherwise the population distribution is a simple Gaussian, and we find $v\approx 1$. Note that this condition is the same as the one we encountered above in the context of whether $C_r(z,v)$ is dominated by young or old clones. For $a_c (1-v)\gg r$, $C_r(z,v)$ and equivalently the normalization of $P(a)$ are dominated by the old clones. The normalization condition on $P(a)$ now requires 
\begin{equation}
\frac{r \sqrt{v}}{(1-v)a_c-r}
e^{\frac{(1-v)a_c^2}{2v}-\frac{ra_c}{v}+\frac{r^2}{2v}}
= 1 \ ,
\end{equation}
which yields
\begin{equation}
\label{eq:norm}
a_c^2 \approx \frac{2v}{(1-v)} \log \frac{(1-v)a_c}{r
\sqrt{v}}\approx  \frac{2v}{(1-v)} \log
\frac{\sqrt{2(1-v)}}{r} \ .
\end{equation}
Solving \EQ{vnose} for $a_c^2$ and equating the result with \EQ{norm} results in
\begin{equation}
\frac{2v}{1-v} \log \frac{\sqrt{2(1-v)}}{r} =
2 \log \frac{\sigma rN}{v} \ .
\end{equation}
After rearranging and replacing $v$ by $\gamma = v/\sigma^2$, we obtain the expression given in \EQ{v_additive}.

\section{Mean field solution}
Above, we discussed the large $r$ solution to the equation
\begin{equation}
\dot P(A,E) = (F-\mfit -r)P(A,E) + \frac{r}{\sqrt{2\pi \sigma_A}}e^{-\frac{(A-\mafit)^2}{2\sigma_A^2}}\rho(E) \ ,
\end{equation}
where we have introduced the symbol $\rho(E)$ for the distribution of epistatic fitness. 
At large $r$, this model admits a factorized solution $P(A,E) =
\vartheta(A,t)\omega(E)$ with 
\begin{equation}
\vartheta(A,t) =
\frac{e^{-\frac{(A-\sigma_At)^2}{2\sigma_A}}}{\sqrt{2\pi\sigma_A^2}} \quad\quad
\omega(E) =
\frac{r\rho(E)}{r+C-E} \ .
\end{equation}
The epistatic part needs both to be normalized, and $C$ needs to equal the mean of the distribution $\omega(E)$. Those two are linked: when one is true, so is the other. Hence
\begin{equation}
\begin{split}
\int dE \frac{r\rho(E)}{r+C-E} &= \int dE \rho(E)\sum_{n=0}^{\infty} \left(\frac{E-C}{r}\right)^n = 1+\int dE \rho(E)\frac{E-C}{r}\sum_{n=0}^{\infty} \left(\frac{E-C}{r}\right)^n \\
& = 1+\frac{1}{r}\int dE E\frac{r\rho(E)}{r+C-E}-\frac{C}{r} = 1+\frac{\mefit-C}{r} =1
\end{split}
\end{equation}
Hence the constant $C$, adjusted to achieve normalization, equals the population mean of $E$.

\end{document}